\documentclass[a4paper]{amsart}

\usepackage{amssymb,amsmath,graphics,epsfig,color}
\usepackage{latexsym}
\usepackage{eucal}
\makeatletter
\@addtoreset{equation}{section}\makeatother

\newtheorem{theo}{Theorem}[section]
\newtheorem{lem}[theo]{Lemma}
\newtheorem{prop}[theo]{Proposition}

\newtheorem{dfn}[theo]{Definition}
\newtheorem{rmk}{Remark}

\newcommand{\norm}[1]{\left\Vert#1\right\Vert}
\newcommand{\beq}{\begin{equation}}
\newcommand{\eeq}{\end{equation}}
\newcommand{\bce}{\begin{center}}
\newcommand{\ece}{\end{center}}
\newcommand{\barr}{\begin{array}}
\newcommand{\earr}{\end{array}}
\newcommand{\ben}{\begin{enumerate}}
\newcommand{\een}{\end{enumerate}}
\newcommand{\li}{\mathcal{L}}
\newcommand{\rr}{\mathbb{R}}
\newcommand{\nn}{\mathbb{N}}
\newcommand{\cc}{\mathbb{C}}
\newcommand{\zz}{\mathbb{Z}}
\newcommand{\E}{\mathcal{E}}
\newcommand{\Le}{\mathcal{L}_E}
\newcommand{\disp}[1]{\displaystyle{#1}}

\newcommand{\rd}{\mathbb{R}^{n}}
\newcommand{\rdd}{\mathbb{R}^{2n}}
\newcommand{\supp}{\mbox{supp}}
\newcommand{\tr}{\mbox{Tr}}
\newcommand{\A}{\mathcal{A}}

\newcommand{\F}{\mathcal{F}}
\newcommand{\Z}{\mathcal{Z}}

\newcommand{\ud}{\frac{1}{2}}

\newcommand{\wt}{\widehat{W_t}}

\newcommand{\tm}{\tilde{M}}

\newcommand{\vect}{\mbox{Span }}
\newcommand{\ima}{\mbox{Im }}

\newcommand{\grad}{\nabla H(z)}
\newcommand{\jgrad}{J\nabla H(z)}

\newcommand{\ce}{\mathcal{C}_{E}}
\newcommand{\ceg}{\mathcal{C}_{E,g}}

\newcommand{\hta}{\widetilde{H}}
\newcommand{\nrja}{\widehat{\Sigma}_E}
\newcommand{\tore}{\mathbb{T}^n}
\newcommand{\alg}{\mathcal{G}}

\newcommand{\gmu}{g^{-1}}
\newcommand{\la}{\lambda}
\newcommand{\e}{\varepsilon}
\newcommand{\aaa}{\alpha}

\newcommand{\nrj}{\Sigma_E}

\newcommand{\vfi}{\varphi}

\newcommand{\lde}{L^2(\mathbb{R}^n)}

\newcommand{\op}{Op_h^w}

\newcommand{\hq}{\widehat{H}}

\newcommand{\gra}{\int_{\mathbb{R} ^{2d}_{\alpha}} }

\begin{document}
\title[Semi-classical trace formula, isochronous case. Application to conservative systems]
{Semi-classical trace formula, isochronous case. Application to conservative systems}
\author[Roch Cassanas]{Roch Cassanas}
\address{Mathematisches Instit\"ut G\"ottingen\\
         Bunsenstr.3-5\\    
         D-37073, G\"ottingen, Germany}
\email{cassanas@uni-math.gwdg.de}
\thanks{This research was financed by the grant RA 1370/2-1 of the German Research Foundation (DFG).\\
Part of this work found its motivation during a short visit grant funded by the ESF program SPECT.}
\subjclass[2000]{Primary 81Q20, Secondary 81R12, Tertiary 81R30}
%
\begin{abstract}
\noindent 
Under conditions of clean flow we compute the leading term in the STF when the set of periods of the energy surface is discrete. Comparing to the case of non-degenerate periodic orbits, we obtain a supplementary term which is given in terms of the linearized flow. As particular cases, we give a STF for quadratic Hamiltonians and we obtain the Berry-Tabor formula for integrable systems. For conservative systems (i.e. systems with several first integrals), we give practical conditions to get a clean flow and interpret the leading term of the STF for a compact symmetry. We give several examples to illustrate our computation. 
\end{abstract}
\maketitle

\section{Introduction}
Let $H:\rdd\to \rr$ be a classical smooth Hamiltonian. Under usual hypotheses (cf (\ref{hypCF})), one can define its Weyl quantization $\hq$ and obtain a selfadjoint operator on $\lde$. Let $E\in\rr$ and $\e>0$. If $H^{-1}([E-\e,E+\e])$ is compact then the spectrum of $\hq$ is discrete in $]E-\e,E+\e[$. To describe the spectrum in  this interval, physicists introduce the local spectral density at $E$ defined by the distribution
\beq\label{spectraldensity}
\mathcal{D}_E(h)=\sum_{j\geq 0}\delta_{\lambda_j(h)},
\eeq
where $\lambda_j(h)$ are the eigenvalues of $\hq$ in $]E-\e,E+\e[$, and they study approximations of $\mathcal{D}_E(h)$. On a mathematical point of view, it is more convenient to study the asymptotics when $h\to 0^+$ of a continuous fonction in $E$, 
\beq\label{densitygutz}
\frac{1}{h}\mathcal{G}_E(h):=\frac{1}{h}\tr\left(\psi(\hq) f\left( \frac{E-\hq}{h} \right) \right)
\eeq
where $\psi$ is a smooth compactly supported function in a neighbourhood of $E$ (energy cut-off) and the Fourier transform of $f$ is also with compact support. The function (\ref{densitygutz}) is the regularization of $\mathcal{D}_E(h)$ in a sense we should describe in section \ref{STFtheo} (cf Lemma \ref{justifdensity}).
Asymptotics of $\mathcal{G}_E(h)$ when $h\to 0^+$ is called the Semi-classical Trace Formula (STF in short). Modulo oscillatory terms, under some "clean flow conditions" (cf definition \ref{def1}) the expansion is a power series in $h$. Its remarkable property lies in the fact that coefficients of this series can be computed in terms of quantities describing periodic orbits of the classical Hamiltonian system associated to $H$ in the energy shell $\nrj:=\{H=E\}$. The most celebrated version is the so called Gutzwiller formula for which one supposes that periodic orbits of $\nrj$ with period in a compact set are in finite number. The leading coefficient can be expressed as a sum over these periodic orbits which involves their primitive period, action, Maslov index and linearized Poincar\'e map (cf (\ref{gutzND})).\

This situation with only few periodic orbits appears for example for mixing flows, as the geodesic flow on a compact manifold of negative curvature.  Nevertheless, in $\rd$, it will never be satisfied when the Hamiltonian system owns some symmetries or first integrals. Indeed, if $H$ is invariant by a one parameter symplectic group, then the image of a periodic orbit of $\nrj$ by an element of this group is also a periodic orbit of $\nrj$ of same period. This proscribes the previous `isolated' status of periodic orbits. Hence, the size of periodic orbit families grows with the symmetries. This is another motivation to investigate such systems, since the bigger periodic orbit manifolds are in $\nrj$, the more they participate to oscillations of $\mathcal{G}_E(h)$ (see the leading power of $h$ in Theorem \ref{gutztheo}). To our knowledge, only very few of such situations with symmetry were studied in this framework, which should be our concern in this paper. Our purpose is to give simple criterions to get a nice description of families of periodic orbits together with an asymptotic expansion of $\mathcal{G}_E(h)$, and also to compute coefficients in such a way to interpret them as geometrical features of the classical motion as far as possible.\

Our article is structured as follows: in section \ref{STFtheo}, under quite general clean flow conditions, we perform a theoretical computation of the leading coefficient of the asymptotics in terms of the linearized flow and the structure of its algebraic eigenspaces (Theorem \ref{gutztheo2}). This is achieved by assuming that the set of concerned periods of $\nrj$ is discrete (`isochronous case'). Our calculations are done pushing further a method based on coherent states due to Combescure, Ralston and Robert (\cite{CRR}), which is to our opinion well fitted to make the linearized flow appear.
 In section \ref{classicalexamples}, we apply this calculation to find back already known cases of STF, as the `Weyl term' (zero period, cf (\ref{weyltermform})), the non-degenerate case (`isolated' periodic orbits, cf (\ref{gutzND})), or the case of periodic flow on $\nrj$(cf (\ref{perflowform})). We also give a STF for quadratic Hamiltonians (Proposition \ref{STFquadrat}). In section \ref{conservative}, we generalize the concept of non-degenerate periodic orbits to conservative systems (i.e. systems with several independant first integrals), via the notion of `normal periodic orbits' already introduced in \cite{Vdb} and \cite{SMK}. We show that it is a natural criterion to obtain a clean flow and a nice description of families of periodic orbits which arise in manifolds (Propositions \ref{normalFP} and \ref{STFgroup}). In section \ref{berrytabor}, we apply this last concept to the integrable case and interpret the leading coefficient in terms of action/angle variables. The STF involves the frequency of periodic tori and the Gaussian curvature of the energy shell in action coordinates (Theorem \ref{STFberrytabor}). This way, we recover the formula of physicists Berry and Tabor (cf \cite{BT1},\cite{BT2}) in a mathematical framework.\\\

\textbf{Aknowledgements}: We thank J. Bolte, B. Camus, A. Laptev, D. Robert and S. V\~u Ng\d oc for stimulating discussions on the subject.
\section{Theoretical STF for isochronous periodic orbits}\label{STFtheo}

\subsection{Statement of the result}
Let $H:\rdd\to \rr$ be a smooth Hamiltonian with associated dynamical system:
\beq\label{hamsyst}
\dot{z}_t=J\nabla H(z_t), \mbox{ where } J=\left(
\barr{cc}
0    & I_n\\
-I_n & 0
\earr\right).
\eeq
We will use the notation $z=(x,\xi)\in\rd\times\rd$ as a variable in $\rdd$. The flow of $H$ at time $t\in \rr$ with initial condition $z\in\rdd$ is denoted by $\Phi_t(z)=(q_t,p_t)$. The trajectory of $z$ under this flow will be denoted by $\gamma_z$. If $E\in\rr$, then $\nrj:=\{H=E\}\subset\rdd$. We define the monodromy matrix or linearized flow as
\beq\label{monodromy}
M_z(t):=\partial_z \Phi_t(z)
\eeq
If $\lambda$ is in the spectrum of $M_z(t)$, we define the algebraic eigenspace $E_\lambda(z,t)$ by
\beq\label{aes}
E_\lambda=E_\lambda(t,z):=\sum_{k=1}^{2n} \ker(M_z(t)-\la Id)^k.
\eeq
As we will be concerned with periodic orbits, the eigenvalue $\lambda=1$ should play a crucial role in the following.\

We make usual hypotheses of quantization on $H$ which allow a nice functional calculus, namely, we suppose that there exists $m>0$ such that
\beq\label{hypCF}
\left\{
\begin{array}{l}
<H(z)>\leq C<H(z^{\prime})><z-z^{\prime}>^m, \quad \forall z,z^{\prime} \in \rr^{2n}.\\
|\partial_z^{\alpha} H(z)|\leq C_{\alpha}<H(z)>, \quad\forall z \in \rr^{2n}, \forall\alpha \in
\nn^{2n}.\\
H \mbox{ has a lower bound on }\rdd.
\end{array}
\right.
\eeq
The Weyl quantization of $H$  is defined as follows: for $u\in \mathcal{S}(\rd)$,
\beq\label{pseudoweyl}
\op(H) u(x)=(2\pi h)^{-n}\int_{\rd}\int_{\rd}e^{\frac{i}{h}(x-y)\xi}
H\left(\frac{x+y}{2},\xi\right) u(y) dy d\xi.
\eeq
In particular, under hypothesis (\ref{hypCF}), $\op(H)$ is essentially selfadjoint on $\mathcal{S}(\rd)$ (see \cite{He-Ro1}), and we denote by $D(\hq)$, $\hq$ its selfadjoint extension on $\lde$. We fix $E\in\rr$. We have in mind to study the spectrum of $\hq$ near the energy $E$. For this purpose, we define the following  {\it regularized spectral density} :
\beq\label{densitespectrale}
\mathcal{G}_E(h):=\tr\left(\psi(\hq) f\left( \frac{E-\hq}{h} \right) \right),
\eeq
where $\psi$ is smooth, compactly supported in a neighbourhood $]E-\e, E+\e[$ of $E$
(where $\e>0$) such that, if $\nrj:=\{ H=E\}\subset \rdd$, then
\beq
\mbox{$H^{-1}([E-\e, E+\e])$ is compact and $\nrj$ has no critical point of $H$.}
\eeq
$\psi(\hq)$ is an energy cut-off which is trace class by  \cite{He-Ro1}.
The function $f$ is such that its Fourier transform $\hat{f}$ is compactly supported in $\rr$.
The justification of definition (\ref{densitespectrale}) is given by the following lemma whose proof is left to the reader:
\begin{lem}\label{justifdensity}
Suppose moreover that $0<\e_1<\e$ and that $\psi=1$ on $[E-\e_1,E+\e_1]$. Then $\psi \mathcal{D}_E(h)=\mathcal{D}_E(h)$ on $\mathcal{D}'(]E-\e_1,E+\e_1[)$ and $(\psi \mathcal{D}_E(h))*f_h=\frac{1}{h}\mathcal{G}_E(h)$, where $f_\tau(t):=\frac{1}{\tau}f(t/\tau)$.
\end{lem}
Thus, if $\psi$ is taken to be constant equal to $1$ in a neighbourhood of $E$, and if $\int_{\rr} f=1$ (or $\hat{f}(0)=1$), then $ \frac{1}{h}\mathcal{G}_E(h)$ is simply a regularization of $\mathcal{D}_E(h)$ via a convolution by an approximate identity. The study of $\mathcal{G}_E(h)$ has proven to be an essential tool to investigate the spectrum of $\hq$. Well known applications are e.g. the asymptotics of the counting function of eigenvalues of $\hq$ in an interval (see \cite{Ro}), or the localization of the spectrum for integrable systems or periodic energy shells by Bohr-Sommerfeld conditions (\cite{Do2}, \cite{AC}, or \cite{CdV2}).\

One condition to get asymptotics of $\mathcal{G}_E(h)$ when $h\to 0$ is to have a {\bf clean flow} on $ \supp \hat{f}\times\nrj$.
Let us introduce the set
$$\mathcal{C}_E:=\{ (t,z)\in \supp \hat{f}\times \nrj : \Phi_t(z)=z \}.$$
\begin{dfn}\label{def1}
We say that the flow is clean on $\supp \hat{f}\times \nrj$ if $\mathcal{C}_E$ is a finite union of submanifolds of $\rr\times\rdd$
and if at each point $(T,z)$ of $\mathcal{C}_E$, we have :
$$\{ (\tau,\aaa)\in \rr\times T_z\nrj:  \tau \jgrad + (M_z(T)-Id)\aaa=0 \}\subset T_{(T,z)}\mathcal{C}_E.$$
\end{dfn} 
When it holds, the last inclusion is actually automatically an equality. If the flow is clean on $\supp \hat{f}\times\nrj$, then $\mathcal{G}_E(h)$
has an asymptotic expansion in powers of $h$ when $h\to 0^+$, possibly multiplied by oscillating terms of the form
$e^{\frac{i}{h}a}$, $a\in\rr$ (see for example \cite{PU2}, \cite{BU}, or \cite{CRR}). There are usual assumptions to ensure that the flow is clean. One can for example suppose that periodic orbits of $\supp \hat{f}\times \nrj$ are non-degenerate. An other one is to suppose that the flow is periodic with same primitive period on $\nrj$ (see section \ref{classicalexamples}). We shall give later other explicit situations of clean flow, in particular for systems with several constants of motion (see section \ref{conservative}).
Let us denote by $[ \ce ]$ the set of connected components of $\ce$. If $(T,z)\in\ce$, then the classical action
\beq\label{action}
\A(T,z):=\int_{0}^T p_t\dot{q_t}dt
\eeq
 is constant on each element of $[ \ce ]$ (see \cite{GS} p.167).
 The following theorem generalizes the results of Colin de Verdi\`ere \cite{CdV}, Chazarain \cite{Ch}, and Duistermaat-Guillemin (\cite{DG}) on compact manifolds. Its proof goes back to the works of Guillemin-Uribe \cite{GU}, Paul-Uribe \cite{PU1} \cite{PU2}, Brummelhuis-Uribe \cite{BU}, Meinrenken \cite{Mein} or Dozias \cite{Do}, who gave mathematical sense to the heuristic first statement of physicists Gutzwiller \cite{Gutzwiller} and Balian-Bloch \cite{balianbloch}. For more details on bibliography, we refer the reader to the nice survey \cite{U}.
\begin{theo}\label{gutztheo}
Suppose that the flow is clean on $\supp{\hat{f}}\times \nrj$. Then we have when $h\to0^+$
$$\mathcal{G}_E(h)=\sum_{Y\in[\ce]}(2\pi h)^{\frac{1-\dim Y}{2}}e^{\frac{i}{h}\A_Y}\frac{\psi(E)}{2\pi} 
\left( \int_Y \hat{f}(t) d(t,z) d\sigma_Y(t,z)+\sum_{j\geq 1}h^j a_{j,Y}\right)+O(h^{\infty}).$$
where $d\sigma_Y$ is the Riemannian measure on $Y$, and
$d(t,z)$ is a density to be specified on $Y$, we shall call the Duistermaat-Guillemin density (cf \cite{DG}), and $\A_Y$ is the common action on $Y$.
\end{theo}
\noindent The Duistermaat-Guillemin density $d(t,z)$ (in short {\it DG-density}) owns lots of the classical quantities characterizing periodic orbits of the Hamiltonian system (\ref{hamsyst}). Note that here it should be considered as a {\it complex} density, contrary to \cite{DG}. This is due to the fact that we should prove our results via coherent states and not using FIO. This method introduced in \cite{CRR} skips the problems of caustics appearing in the usual WKB method, but requires complex phases, which explains the fact that our DG-density is also complex. Note that, to our point of view, the use of coherent states makes the computations more explicit in terms of the linearized flow $M_z(T)$ than the one given by FIO method (see Remark \ref{rmk1} of section \ref{Proofgutz}).  As a consequence, our DG-density should include Maslov indices (see section \ref{classicalexamples}). Its computation is in general non-trivial and, to our knowledge, has only been achieved in some particular cases of importance (see section \ref{classicalexamples}). It usually appears as the determinant of a transversal Hessian coming from a stationary phase theorem. In \cite{DG}, Lemma 4.4, some computation of $d$ is achieved in terms of  half densities, when one has:
\beq\label{hypDG}
\forall (T,z)\in\mathcal{C}_E, \; \ker(M_z(T)-Id)^2\cap T_z\nrj=\ker(M_z(T)-Id)\cap T_z\nrj.
\eeq
This hypothesis takes into account the cases of the period $T=0$ (Weyl term), the case of non-degenerate periodic orbits or the case of periodical flow on $\nrj$. However, when one considers systems with constants of motion in involution, this assumption is no more fulfilled as we shall see later (cf Lemma \ref{Meyer}). Our main motivation in this section is to broaden the computation to more general settings. Let us denote the set of periods by
\beq
\mathcal{L}_E:=\{T\in\supp \hat{f}: \exists z\in \nrj, \Phi_T(z)=z\}.
\eeq
For $T\in \mathcal{L}_E$, let
$$\mathcal{Z}_T:=\{z\in\nrj:\Phi_{T}(z)=z\}.$$

As in \cite{DG}, we shall compute the DG-densities in the '{\it isochronous case}', i.e. when $\mathcal{L}_E$ is finite.
\begin{theo}\label{gutztheo2}
Suppose that
\beq\label{LEfinite}
\mbox{$\mathcal{L}_E$ is finite}
\eeq
and that for all $T$ in $\mathcal{L}_E$, $\mathcal{Z}_T$ is a finite union of smooth manifolds such that
$\forall z\in\mathcal{Z}_T$
\beq\label{tangent}
T_z \mathcal{Z}_T=\ker(M_z(T)-I)\cap T_z\nrj \mbox{ and } \jgrad\notin(M_z(T)-I)(T_z\nrj).
\eeq
Then the flow is clean on $\supp\hat{f}\times\nrj$, and we have modulo $O(h^{+\infty})$ when $h\to 0^+$:
\beq
\mathcal{G}_E(h)=\psi(E)\sum_{T\in\Le}\sum_{Y\in[\Z_T]}(2\pi h)^{\frac{1-\dim Y}{2}}e^{\frac{i}{h}\A(T,Y)}\hat{f}(T)\frac{1}{2\pi}
\left( \int_Y  d(T,z) d\sigma_Y(z)+\sum_{j\geq 1}h^j a_{j,Y}\right),
\eeq
 where $[\Z_T]$ is the set of connected components of $\Z_T$, $d\sigma_Y$ is the Riemannian measure on $Y$, and $\A(T,Y)$ is the constant value of the action $\A$ on $\{T\}\times Y$.\

Moreover if we suppose that
\beq\label{hypRC}
\forall (T,z)\in\mathcal{C}_E, \; E_1(T,z)=\ker(M_z(T)-Id)^2.
\eeq
then we have in addition that for all $(T,z)$ in $\ce$
\beq
d(T,z)^2=\frac{(-1)^n i^{-(k+1)}\det({w_0}_{|_{E_1}})}{\det(M_z(T)-I)_{|_{V_1}} \norm{\Pi_{E_1}(\grad)}^2 \det(\Pi_{\E_5} J(M_z(T)-I)_{|_{\E_5}} ) }.
\eeq
where $w_0(.,.):=<J., .>_{\rdd}$ is the usual symplectic form on $\rdd$, $V_1=(JE_1(T,z))^\perp$, and $\Pi_{E_1}$ is the orthogonal projection on $E_1(T,z)$. If $\E_1:=\ker(M_z(T)-I)\cap T_z\nrj$, then $k:=\dim\E_1$, $\E_5:=\E_1^{\perp}\cap(E_1\cap T_z\nrj)$, and $\Pi_{\E_5}$ is the orthogonal projection on $\E_5(T,z)$.
\end{theo}
\noindent{\it Remarks}
\begin{itemize}
\item The determinant $\det(M_z(T)-I)_{|_{V_1}}$ is non zero since, $M_z(T)$ being a symplectic map, $V_1(T,z)$ is the sum of all algebraic eigenspaces of $M_z(T)$ corresponding to eigenvalues different from $1$.
\item Note that there remains an ambiguity to obtain $d(T,z)$ from $d(T,z)^2$. This will give entire powers of $e^{i\frac{\pi}{4}}$ we should compute in particular cases, and this should involve Maslov indices of periodic orbits of $\mathcal{Z}_T$ (see \cite{MDS}).
\item We generalize the computations of \cite{DG} in the sense that (\ref{hypDG}) implies (\ref{hypRC}) and $\jgrad\notin (M_z(T)-Id)(T_z\nrj)$. \footnote{Indeed, assuming (\ref{hypDG}), if $x\in\rdd$ is such that $(M_z(T)-Id)^3x=0$, then $(M_z(T)-Id)x\in T_z\nrj\cap\ker(M_z(T)-Id)^2=T_z\nrj\cap\ker(M_z(T)-Id)$. Thus $(M_z(T)-Id)^2x=0$. This yields (\ref{hypRC}).
Moreover, if $\jgrad=(M_z(T)-Id)y$, with $y\in T_z\nrj$, then $y\in \ker(M_z(T)-Id)^2\cap T_z\nrj=\ker(M_z(T)-Id)\cap T_z\nrj$, thus $\jgrad=0$, which is excluded.} Under these supplementary assumptions in Theorem \ref{gutztheo}, we have $\E_5=\{0\}$ and 
\beq\label{DGsimple}
d(T,z)^2=\frac{(-1)^{n+\frac{\dim E_1}{2}} \det({w_0}_{|_{E_1}})}{\det(M_z(T)-I)_{|_{V_1}} \norm{\Pi_{E_1}(\grad)}^2}.
\eeq
\end{itemize}

\subsection{Proof of Theorem \ref{gutztheo2}}\label{Proofgutz}
We use the method employed in \cite{CRR} based on coherent states. We refer to \cite{Cas1} for the following reduction\footnote{Take $G=\{Id \}$ in this case.}(equations (3.15) and (3.17) of this reference). When $h\to0^+$, the trace $\mathcal{G}_E(h)$ has the same behaviour in leading order than the following integral
\beq\label{proof1}
I_E(h):=\frac{(2\pi h)^{-d}}{2\pi}\int_{\rr_t}\gra \exp\left(\frac{i}{h}\varphi_{E}(t,z)\right) \frac{\hat{f}(t)
\psi(H(z))}{ \det_*^{\ud}\left( \frac{A_t+iB_t-i(C_t+iD_t)}{2}\right)} dtdz.
\eeq
where $A_t,B_t,C_t,D_t$ are the four $n\times n$ block matrices of
$$M_z(t)=\left(
\begin{array}{cc}
A_t & B_t\\
C_t & D_t
\end{array}\right),$$
and $\varphi_{E}:\rr\times\rdd\to \cc$ is the {\it complex} phase given by
$$\vfi_E=\vfi_1+i \vfi_2.$$
\beq\label{proof2}
\left\{
\barr{l}
\disp{ \vfi_1(t,z):=(E-H(z))t-\ud\int_0^t(z_s-z)J\dot{z_s}ds}\\
\disp{ \vfi_2(t,z):=\frac{i}{4}<(I-\wt)(z_t-z);(z_t-z)>.}
\earr\right.
\eeq
where
$\widehat{W_t}:=\left(
\barr{cc}
W_t      & -iW_t\\
-i\, W_t & -W_t
\earr
\right)$
with $\ud (I+W_t):=(I-i^t\, M_0)^{-1}$ and $M_0:=(C_t+iD_t)(A_t+iB_t)^{-1}$.
Moreover, we have:
\beq\label{proof3}
\norm{W_t}_{\li(\cc^{d})}<1.
\eeq
We send the reader to \cite{CRR} (Theorem 3.3) for the precise meaning of $\det_*^{\ud}$. We just recall that
$$[\mbox{det}_*^{\ud}]^2=\det.$$
It is shown in \cite{Cas1} (Proposition 4.1) that if $\ce:=\{\vfi_E'=0\}\cap\{\Im \vfi_E=0\}\cap(\supp\hat{f}\times\rdd)$, then one has
\beq\label{proof4}
\ce=\{ (t,z)\in \supp \hat{f}\times \nrj : \Phi_t(z)=z \}.
\eeq
Moreover, an integration by part shows that
\beq
\forall (T,z)\in \ce, \quad \vfi_E(T,z)=\A(T,z).
\eeq
\begin{rmk}\label{rmk1}
Let us compare with the usual WKB method using FIO. In this case, one boils down to the same kind of integral as in (\ref{proof1}), but with the {\it real} phase
$$\tilde{\vfi}_E(t,z)=S(t,z)-x.\xi+tE,$$
where $S$ satisfies the Hamilton-Jacobi equation
\beq\label{hamjac}
\left\{
\begin{array}{l}
\partial_t S(t,z)=-H(x,\nabla_x S(t,z)).\\
S(0,z)=x.\xi.
\end{array}\right.
\eeq
Actually our $\A(t,z)$ is just the constant value of $S(t,z)-x.\xi+tE$ on $\ce$ (see e.g. \cite{PU2}, \cite{U}). It appears that for the computations of the leading term of the asymptotics of $\mathcal{G}_E(h)$, the phase $\vfi_E$ issued from coherent states seems to us  more explicit in terms of the linearized flow $M_z(t)$ than the phase $\tilde{\vfi}_E$ issued from FIO and defined implicitly via the equation (\ref{hamjac}).
\end{rmk}
The conditions under which one can apply the generalized stationary phase theorem (see \cite{CRR} Theorem 3.3) to integral (\ref{proof1}) are exactly the clean flow conditions given by Definition \ref{def1} (see \cite{Cas1} Proposition 4.3). Writing 
$$\ce=\underset{T\in\mathcal{L}_E}{\bigcup}\{T\}\times\Z_T,$$
it is easy to see that under hypotheses (\ref{LEfinite}) and (\ref{tangent}), the flow is clean on $(\supp\hat{f})\times\nrj$. Indeed in our case, if $T\in \supp\hat{f}$ and $\Phi_T(z)=z\in\nrj$, then $T_{(T,z)}\ce=\{0\}\times T_z\Z_T$. Applying the stationary phase theorem yields modulo $O(h^{\infty})$ when $h\to 0^+$:
\beq
\mathcal{G}_E(h)=\psi(E)\sum_{T\in\Le}\sum_{Y\in[\Z_T]}(2\pi h)^{\frac{1-\dim Y}{2}}e^{\frac{i}{h}\A(T,Y)}\hat{f}(T)\frac{1}{2\pi} 
\left( \int_Y  d(T,z) d\sigma_Y(z)+\sum_{j\geq 1}h^j a_{j,Y}\right).
\eeq
with
\beq\label{proof5}
d(T,z):=\mbox{det}_*^{-\ud}\left( \frac{\varphi_{E}^{\prime\prime}(T,z)_{|_{\mathcal{N}_{(T,z)}\ce}}}{i} \right)
\mbox{det}_*^{-\ud}\left( \frac{A_T+iB_T-i(C_T+iD_T)}{2}\right).
\eeq
where $[\Z_T]$ is the set of connected components of $\Z_T$ , $d\sigma_Y$ is the Riemannian measure on $Y$, and
$\A(T,Y)$ is the constant value of the action $\A$ on $\{T\}\times Y$.
We have 
$$\det\left( \frac{\varphi_{E}^{\prime\prime}(T,z)_{|_{\mathcal{N}_{(T,z)}\ce}}}{i} \right)=
\det\left(\frac{\mbox{Hess } \vfi_E(T,z)+i\Pi_{T_{(T,z)}\ce}}{i}\right).$$
where $\Pi_{T_{(T,z)}\ce}$ is the orthogonal projection on $T_{(T,z)}\ce$ and $\mbox{Hess } \vfi_E(T,z)$ is given by \cite{Cas1} Proposition 4.2 (take $g=Id$ in our case). This implies that
$\det\left( \frac{\varphi_{E}^{\prime\prime}(T,z)_{|_{\mathcal{N}_{(T,z)}\ce}}}{i} \right)=$
$$\det\left(
\barr{c|c}
\frac{1}{2}<(I-\widehat{W}_T) \jgrad;\jgrad> & -\frac{1}{i} ^t\grad        \\
 & + \frac{1}{2}^t\left[(^tM_z(T)-I)(I-\widehat{W}_T) \jgrad \right] \\\hline

-\frac{1}{i}\grad & \frac{1}{2i}[JM_z(T) +^t(JM_z(T))] \\
+\frac{1}{2}(^tM_z(T)-I)(I-\widehat{W}_T) \jgrad & +\frac{1}{2}(^tM_z(T)-I)(I-\widehat{W}_T)(M_z(T)-I)+\Pi_{\E_1}\\
\earr
\right),$$
where $\Pi_{\E_1}$ is the orthogonal projection on $\E_1:=T_z\Z_T$.
Since $M_z(T)$ is symplectic, we have 
$$JM_z(T)+^t(JM_z(T))=(^tM_z(T)+I)J(M_z(T)-I).$$
Set:
\beq
K:=\frac{1}{2i} (^tM_z(T)+I)J+\ud (^tM_z(T)-I)(I-\widehat{W}_T).
\eeq
Then, the fourth block is equal to $K(M_z(T)-I)+\Pi_{\E_1}$.\\
We recall that, classically, $\jgrad \in \ker(M_z(T)-Id)$ and thus $\grad\in\ker( ^tM_z(T)-Id)$. Using this, we note that the third block is equal to $K\jgrad$.
Let us set:
\beq
X_1:=\ud (I-\widehat{W}_T)\jgrad.
\eeq
We get:
$$\det\left( \frac{\varphi_{E}^{\prime\prime}(T,z)_{|_{\mathcal{N}_{(T,z)}\ceg}}}{i} \right)=
\det\left(
\barr{c|c}
^tX_1\jgrad & i^t\grad +^tX_1(M_z(T)-I)\\\hline
K\jgrad     & K(M_z(T)-I) +\Pi_{\E_1}
\earr\right).$$
Now, in view of \cite{Cas1}, Lemma 4.10 and our equation (\ref{proof5}), we have
$$d(T,z)^{-2}=(-1)^n\det\left(\barr{c|c}
^tX_1\jgrad & i^t\grad +^tX_1(M_z(T)-I)\\\hline
\jgrad     & (M_z(T)-I) +K^{-1}\Pi_{\E_1}
\earr
\right).$$
Moreover\footnote{Note that there is a slight misprint in \cite{Cas1} at this point, one minus sign is missing.}
\beq\label{proof6}
K^{-1}=-\ud [(M_z(T)-I)+i(M_z(T)+I)J].
\eeq
We denote by
\beq
\aaa:=<X_1,\jgrad>.
\eeq
Note that $\aaa\neq 0$ since for $\alpha=(q,p)$, we have (\ref{proof4}), $<\widehat{W}_T\alpha,\alpha>=<W_T(q-ip),(q-ip)>$, and $\jgrad\neq 0$.
We use the line operation $L_2\gets L_2-\frac{1}{\aaa} \jgrad L_1$, to get:
\beq\label{proof7}
d(T,z)^{-2}=(-1)^n \aaa \det(D).
\eeq
where
$$D:=(M_z(T)-I) +K^{-1}\Pi_{\E_1}-\frac{1}{\aaa}\jgrad [i^t\grad +^tX_1(M_z(T)-I)].$$
Now, in view of (\ref{tangent}), we have
$K^{-1}\Pi_{\E_1}=\frac{i}{2}(M_z(T)+I)J\Pi_{\E_1}$ and therefore
\beq\label{proof8}
D=(M_z(T)-I) -\frac{i}{2}(M_z(T)+I)J\Pi_{\E_1}-\frac{1}{\aaa}\jgrad [i^t\grad +^tX_1(M_z(T)-I)].
\eeq
We are now going to compute the determinant of $D$ in a basis of $\rdd$ fitted to the linearized flow $M_z(T)$. To do this we need a supplementary hypothesis describing $E_1(T,z)$. Let us suppose that we have (\ref{hypRC}).
\begin{lem}\label{Meyer}\cite{Meyer}
Let $M$ be a symplectic matrix and note $E_1:=\sum_{k\geq 1}\ker(M-Id)^k$. Suppose that there exist two vector spaces $V$ and $W$ such that
$$W\subset V\subset E_1.$$
$$\forall (x,y)\in W\times V,\quad w_0(x,y)=0.$$
Then $\dim E_1\geq \dim V +\dim W$.
\end{lem}
\begin{proof}
We recall that, if $V_1:=(JE_1)^\perp$, then $V_1$ is the sum of all algebraic vector spaces of $M$ corresponding to eigenvalues different from one, and so we have $V_1\cap E_1=\{0\}$. In particular, we get $W\cap V_1=\{0\}$, or equivalently, $JW\cap E_1^\perp =\{0\}$. But, by hypothesis, we have $JW\subset V^\perp$ and $E_1^\perp\subset V^\perp$. Thus
$$\dim V^\perp\geq \dim E_1^\perp + \dim W,$$
which proves the result.
\end{proof}
\begin{lem}
Let $\E_2=\E_2(T,z):=\{x\in\E_1:\forall y\in \E_1, w_0(x,y)=0 \}.$
We have \
$$E_1=\ker(M_z(T)-Id)^2\iff \dim E_1=\dim \E_1+\dim\E_2.$$
\end{lem}
\begin{proof}
Let us suppose that $\dim E_1=\dim \E_1+\dim\E_2$. Coming back to the proof of Lemma \ref{Meyer} with $W=\E_2$ and $V=\E_1$, we have for dimensional reasons 
\beq\label{proof9}
J\E_2\oplus E_1^\perp=\E_1^\perp.
\eeq
We recall that, by a symplectic argument, $\ker( ^tM_z(T)-Id)=J\ker(M_z(T)-Id)$. Thus we have
$(M_z(T)-Id)(E_1)\subset (J\E_1)^\perp\cap E_1$. Together with (\ref{proof9}), this implies that
$$(M_z(T)-Id)(E_1)\subset (\E_2+(JE_1)^\perp)\cap E_1=\E_2.$$
Thus $E_1=\ker(M_z(T)-Id)^2$.\

Reciprocally, let us suppose that $E_1=\ker(M_z(T)-Id)^2$. By Lemma \ref{Meyer}, we have $\dim E_1\geq\dim \E_1+\dim\E_2$. Let us prove the inverse inequality. Let $\F$ be a supplementary vector space of $\E_1$ in $E_1$. One has to show that $\dim \F\leq \dim \E_2$. One of course have that $(M_z(T)-Id)(\F)+\rr\jgrad\subset \E_2$. Thus
\beq\label{proof10}
\dim[(M_z(T)-Id)(\F)+\rr\jgrad]\leq \dim \E_2.
\eeq
If $\ker(M_z(T)-Id)\subset \nrj$, then $\E_1=\ker(M_z(T)-Id)$ and $\dim(M_z(T)-Id)(\F)=\dim\F$. In view of the last equation, we have $\dim \F\leq \dim \E_2$.\

Now if $\ker(M_z(T)-Id)$ is not included in $\nrj$, then equivalently, $\jgrad\notin\ima(M_z(T)-Id)$. Thus 
\beq\label{proof11}
\dim[(M_z(T)-Id)(\F)+\rr\jgrad]=\dim(M_z(T)-Id)(\F) +1.
\eeq
But, $\nrj$ being an hypersurface of $\rdd$, there is at most one element of $\ker(M_z(T)-Id)$ in $\F$. Thus $\dim\F\leq \dim (M_z(T)-Id)(\F) +1$. Together with (\ref{proof10}) and (\ref{proof11}), we get $\dim \F\leq \dim \E_2$, whcih proves our Lemma.
\end{proof}
We recall that we have $\rdd=E_1\oplus V_1$, where $V_1=(JE_1)^\perp$ and that $E_1$ and $V_1$ are invariant by $M_z(T)$. We denote by
\beq
k:=\dim \E_1,\qquad r:=\dim\E_2.
\eeq
\begin{lem}\label{e1}
There exists $\e_1\neq 0$ in $E_1\setminus T_z\nrj$ such that $(M_z(T)-Id)(\e_1)\in\rr\jgrad$.
\end{lem}
\begin{proof}
If $\ker(M_z(T)-Id)\subset T_z\nrj$, we have equivalently $\jgrad\in\ima(M_z(T)-Id)$. The hypothesis (\ref{tangent}) ensures that $\jgrad\notin (M_z(T)-Id)(T_z \nrj)$, and we thus obtain $\e_1$. If $\ker(M_z(T)-Id)$ is not included in $T_z\nrj$, then we take $\e_1$ in $\ker(M_z(T)-Id)\setminus T_z\nrj$.
\end{proof}
We construct a basis of $E_1$ as follows. Let
\begin{itemize}
\item $\alpha_1:=\jgrad$.
\item $\alpha_2,\dots,\alpha_r$ be a basis of $\E_2$
\item $\alpha_{r+1},\dots,\alpha_k$  be a supplementary basis of $\vect(\alpha_1,\dots,\alpha_r)$  in $\E_1$.
\item $\e_1$ be constructed as in Lemma \ref{e1}
\item $\e_2,\dots,\e_r$ be a supplementary basis of  $\vect(\alpha_1,\dots,\alpha_k,\e_1)$ in $E_1$ included in $T_z    \nrj$ .
\end{itemize}
For $j=1,\dots,k$, we have
$$D(\alpha_j)=-\frac{i}{2}(M_z(T)+I)J(\alpha_j).$$
Note that $V_1\subset T_z\nrj$ since $\jgrad\in E_1$. If $v\in V_1$, then
$$D(v)=(M_z(T)-Id)(v)-\frac{1}{\alpha}<X_1,(M_z(T)-Id)(v)>\jgrad -\frac{i}{2}(M_z(T)+Id)J\Pi_{\E_1}(v).$$
If $j=2,\dots,k$, then $D(\e_j)=$
$$(M_z(T)-Id)(\e_j)-\frac{1}{\alpha}[<X_1,(M_z(T)-Id)(\e_j)>+i<\grad,\e_j>]\jgrad -\frac{i}{2}(M_z(T)+Id)J\Pi_{\E_1}(\e_j).$$
$$D(\e_1)=-\frac{i}{2}(M_z(T)+Id)J\Pi_{\E_1}(\e_1) -\frac{i}{\alpha}<\grad,\e_1>\jgrad.$$
Now we can write the matrix of $D$ in the preceeding basis of $\rdd$. Note that since all $D(\alpha_j)$ ($j=1,\dots,k$) span the image of $\frac{i}{2}(M_z(T)+I)J\Pi_{\E_1}$, by column operations, we can remove its contribution to all other columns. Then, since  $<\grad,\e_1>\neq 0$, by substracting multiples of the column of $D(\e_1)$ to other columns, we obtain a first line with only one non zero entry (the one of $D(\e_1)$). Note also that $(M_z(T)-Id)(\e_j)\in \E_2$, thus it can be expressed only in terms of $\alpha_1,\dots,\alpha_r$. Endly, we recall that $(M_z(T)-Id)(v)\in V_1$. We use the following notations
\beq\label{proof12}
\frac{i}{2}(M_z(T)+Id)J(\alpha_j)=\sum_{i=1}^k w_i^j \alpha_i +\sum_{i=1}^r w_{k+i}^j \e_i +w_V^j,
\eeq
where $w_V^j\in V_1$.
\beq\label{proof12bis}
(M_z(T)-Id)(\e_j)=\sum_{i=1}^r \lambda_j^i \alpha_i.
\eeq
\beq
\mathbb{A}:=((w_i^j))_{r+1\leq i,j\leq k+r},\qquad
\mathbb{B}:=((\lambda_j^i))_{2\leq i,j\leq r}.
\eeq
Then using all preceeding remarks, we obtain
\beq\label{proof12ter}
\det(D)=-\frac{i}{\alpha}<\grad,\e_1>\det(M_z(T)-Id)_{|_{V_1}}(-1)^{kr}(-1)^k\det(\mathbb{A})\det(\mathbb{B}).
\eeq
Taking the bracket of equation (\ref{proof12}) with $J\alpha_p$, for $j,p=1,\dots,k$, we obtain
\beq\label{proof13}
i<\alpha_j,\alpha_p>=\sum_{i=r+1}^k w_i^j <\alpha_i,J\alpha_p>+\sum_{i=1}^r w_{k+i}^j <\e_i,J\alpha_p>.
\eeq
Let $X$ and $Y$ be the $k\times k$ matrices defined by
\beq
Y:=((i<\alpha_j,\alpha_p>))_{1\leq p,j\leq k}
\eeq
\beq
X:=\left(
\begin{array}{ccc|ccc}
<\alpha_{r+1},J\alpha_1>& \dots &<\alpha_{k},J\alpha_1> & <\e_{1},J\alpha_1> & \dots & <\e_{r},J\alpha_1>\\
                        & \ddots &                       & &\ddots &\\
<\alpha_{r+1},J\alpha_k>& \dots &<\alpha_{k},J\alpha_k> & <\e_{1},J\alpha_k> & \dots & <\e_{r},J\alpha_k>
\end{array}\right).
\eeq
Then (\ref{proof13}) is equivalent to
$$Y=X\mathbb{A}.$$
Using the definition of $\E_2$ and the fact that $<\e_j,J\alpha_1>=0$ for $j=2,\dots,r$, we get that
$$\det(X)=(-1)^{r(k-r)+1}<\e_1,\grad>\det(<\alpha_i,J\alpha_j>)_{r+1\leq i,j \leq k}\, \det(<\e_i,J\alpha_j>)_{2\leq i,j \leq r}.$$
We obtain, 
$$\alpha\det(D)=(-1)^k\frac{\det(M_z(T)-Id)_{|_{V_1}}(-1)^{kr+1}i^{k+1}\det(\mathbb{B})\det(<\alpha_j,\alpha_p>)_{1\leq j,p\leq k}}
{\det(<J\alpha_i,\e_j>)_{2\leq i,j \leq r}(-1)^{r(k-r)+1}\det(<\alpha_i,J\alpha_j>)_{r+1\leq i,j \leq k} }.$$
Using (\ref{proof12ter}), this means\footnote{Note that, since $k+r=\dim E_1$, which is even, we have $(-1)^{r+k}=1$.}
\beq\label{Creagh}
d(T,z)^2=\frac{(-1)^{n}}{i^{k+1}} \frac{ \det(<J\alpha_i,\e_j>)_{2\leq i,j \leq r}\det(<\alpha_i,J\alpha_j>)_{r+1\leq i,j \leq k}}{\det(M_z(T)-Id)_{|_{V_1}}\det(\mathbb{B})\det(<\alpha_i,\alpha_j>)_{1\leq i,j\leq k}}.
\eeq
Taking the bracket of equation (\ref{proof12bis}) with $-J\e_p$, $p=2,\dots,k$, we obtain
\beq\label{proof14}
\det(<J(M_z(T)-Id)(\e_i),\e_j>)_{2\leq i,j \leq r}=\det(\mathbb{B})\det(<J\alpha_i,\e_j>)_{2\leq i,j \leq r}.
\eeq
\begin{lem}\label{calculdetw0} If $\alpha_2,\dots,\alpha_k$ is taken orthonormal and if $\e_2,\dots,\e_r$ is taken orthonormal and normal to $\E_1$, then
$$\det({w_0}_{|_{E_1}})= \det(<J\alpha_i,\alpha_j>)_{r+1\leq i,j \leq k}\left[\frac{\norm{\Pi_{E_1}(\grad)}}{\norm{\grad}}\det(<\alpha_i,J\e_j>)_{2\leq i,j \leq r}\right]^2.$$
\end{lem}
\begin{proof}
Compute the determinant in the orthonormal basis $(\tilde{\alpha}_1,\alpha_2,\dots,\alpha_k,\tilde{\e}_1,\e_2,\dots,\e_r)$, where $\tilde{\alpha}_1:=\jgrad/\norm{\jgrad}$ and $\tilde{\e}_1:=\Pi_{E_1}(\grad)/\norm{\Pi_{E_1}(\grad)}$.
\end{proof}
Using (\ref{proof14}) and (\ref{Creagh}), we obtain, under the assumptions of Lemma \ref{calculdetw0},
$$d(T,z)^2=\frac{(-1)^{n}}{i^{k+1}} \frac{\det({w_0}_{|_{E_1}})}{\det(M_z(T)-Id)_{|_{V_1}}\det(<J(M_z(T)-Id)(\e_j),\e_i>)_{2\leq i,j \leq r}\norm{\Pi_{E_1}(\grad)}^2}.$$
This ends the proof of Theorem \ref{gutztheo2}.
\section{Classical examples}\label{classicalexamples}
Let us briefly illustrate Theorem \ref{gutztheo2} for the most usual cases. In this section we only deal with examples such that (\ref{hypDG}) is satisfied. We let more general cases for sections \ref{conservative} and \ref{berrytabor}. Thus, one can use formula (\ref{DGsimple}). Of course, by a partition of unity, in the isochronous case, we can always suppose that $\supp\hat{f}\cap \mathcal{L}_E$ is reduced to a single point:
\subsection{The Weyl term}\label{weylterm}
Suppose that $\supp\hat{f}\cap\mathcal{L}_{E} =\{ 0 \}$. Then locally $\ce=Y=\{ 0 \}\times\nrj$. As $M_z(0)=Id$, we have $E_1=\rdd$, $\E_1(0,z)=T_z\nrj$, $\det({w_0}_{|_{E_1}})=1$, $\Pi_{E_1}(\grad)=\grad$, and $V_1=\{ 0 \}$. Therefore
$$d(0,z)^2=\frac{1}{\norm{\nabla H(z)}^2}.$$
This case leads to asymptotics of the counting function of eigenvalues of $\hq$ in a given interval of $\rr$ as $h$ goes to zero (see \cite{Ro}). Theoretically, one can also compute all the terms $a_{j,Y}$ of Theorem \ref{gutztheo} using the asymptotics of $\tr(\vfi(\hq))$ for $\vfi\in C^\infty_c(]E-\e,E+\e[)$ described in \cite{He-Ro1} ("weak asymptotics"). It allows also to solve the ambiguity on the sign of $d(0,z)$. We find in this case
\beq\label{weyltermform}
\mathcal{G}_E(h)=\psi(E)(2\pi h)^{-n+1}\hat{f}(0)\frac{1}{2\pi}\int_{\nrj} \frac{d\sigma_{\nrj}}{\norm{\nabla H}}\; + O(h^{-n+2}),
\eeq
where $d\sigma_{\nrj}$ is the Riemannian measure on $\nrj$.
\subsection{Non degenerate periodic orbits}
Let us recall that for a general Hamiltonian system of the type (\ref{hamsyst}),  a periodic orbit $(T,z)\in \rr^*\times\rdd$ (i.e. such that $\Phi_T(z)=z$), is {\it non-degenerate} if we have $\dim E_1(T,z)=2$.
In a more geometrical way, $1$ is not an eigenvalue of the energy reduced linearized Poincar\'e map. This type of dynamics for periodic orbits arise in particular for the so-called `mixing' systems (see the survey \cite{DB}). One consequence is that $\{T\}\times \gamma_z$ is an isolated connected component of $\ce$ (cf \cite{MH} 'cylinder theorem' p.136 Theorem 10), locally $\ce=\{ T \}\times \gamma_z$. Since $\Pi_{E_1}(\grad)$ is orthogonal to $\jgrad$, we obtain an orthonormal basis of $E_1$ and easily compute $\det({w_0}_{|_{E_1}})=\frac{\norm{\Pi_{E_1}(\grad)}^2}{\norm{\grad}^2}$.
$$d(T,z)^2=\frac{(-1)^{n+1}}{\det(M_z(T)-Id)_{ |_{V_1}} \,\norm{\nabla H(z)}^2}.$$
Passing to $d(T,z)$, one makes the Maslov index of the orbit appear, and after integration on the periodic orbit, this leads to a rigorous Gutzwiller trace formula. Let us suppose that all periodic orbits of $\supp\hat{f}\times\nrj$ are non-degenerate. For sake of simplicity we suppose also that $0\notin \supp\hat{f}$, in order to skip terms of zero period (cf (\ref{weylterm})). Then one has when $h\to 0^+$
\beq\label{gutzND}
\mathcal{G}_E(h)=\sum_{\gamma \mbox{ {\scriptsize P.O. of }}\nrj }\sum_{n\in \zz^*}\psi(E)\hat{f}(nT_\gamma^*)e^{\frac{i}{h}n\A(\gamma)}\frac{T_\gamma^* e^{i\frac{\pi}{2}n\sigma_{\gamma}}}{2\pi \sqrt{|\det((dP)^n-Id)|}} \; + O(h),
\eeq
where $T_\gamma^*$ denotes the primitive period of $\gamma$, $\A(\gamma)$ is the classical action of $\gamma$ ($\A(\gamma)=\int_\gamma pdq$), $\sigma_{\gamma}$ is an integer called the Maslov index of $\gamma$ (see \cite{MDS}), and  $dP$ is the differential of the Poincar\'e map of $\gamma$ restricted to its energy surface.
\subsection{Periodic flow}
(with unique primitive period). Suppose that $T\in\rr^*$ is such that locally $\mathcal{C}_E=\{T\}\times\nrj$. Differentiating $\Phi_T(z)=z$ on $\nrj$, we obtain that $M_z(T)=Id$ on $T_z \nrj$, and $T_z\nrj\subset E_1$. As $E_1$ is even dimensional ($M_z(T)$ is symplectic), this yields $E_1=\rdd$. Moreover $\E_1=T_z \nrj$. Thus
$$d(T,z)^2=\frac{1}{\norm{\nabla H(z)}^2}.$$
This allows in particular to localize the spectrum in this setting, which yields Bohr-Sommerfeld conditions (see \cite{Do2}). If all points of $\nrj$ have same primitive period $T^*$ then we have, without restriction on $\supp\hat{f}$,
\beq\label{perflowform}
\mathcal{G}_E(h)=\psi(E)(2\pi h)^{-n+1}\sum_{p\in\zz}\hat{f}(pT^*)e^{\frac{i}{h}p \A_{E}}e^{i\frac{\pi}{2}p\sigma_E}\frac{1}{2\pi}\int_{\nrj} \frac{d\sigma_{\nrj}}{\norm{\nabla H}}\; + O(h^{-n+2}),
\eeq
where $\A_E$ is the classical action of any orbit of $\nrj$ ($\A_E=\int_\gamma pdq$ if $\gamma$ is an orbit of $\nrj$), $\sigma_{E}$ is their common Maslov index, and  $d\sigma_{\nrj}$ is the Riemannian measure on $\nrj$.
\subsection{Quadratic Hamiltonians}
In this subsection, our aim is to show that already for the simple class of quadratic Hamiltonians, periodic orbits can appear in families of arbitrary dimension and that this gives more `exotic' STF. Here we restrict ourselves to the case where $H$ is given by:
\beq\label{Hw}
H_w(x,\xi):=\ud \, (|\xi|^2+<Sx,x>_{\rd}),
\eeq
where $S$ is a symmetric positive definite $n\times n$ real matrix. As $S$ can be diagonalized in orthonormal basis, it is straightforward to see that we can boil down to the case where $S$ is diagonal, namely $S=diag(w_1^2, \dots,w_n^2)$ where $w=(w_1,\dots,w_n)\in (\rr_+^*)^n$.
Note that this system is "integrable almost everywhere" in the sense that the functions
$$F_i(z):=\ud(\xi_i^2+w_i^2x_i^2), \quad i=1,\dots,n$$
are first integrals of  the system which are in involution, but whose gradients are not linearly independant everywhere. Indeed, if $z=(x,\xi)\in\rdd$, whenever with have simultaneously $x_j=\xi_j=0$, the gradient of $F_j$ vanishes at $z$ and thus the gradients are colinear. We shall see that this situation precisely appears where the orbits are periodic!
As a consequence, in a neighbourhood of these points, the dynamics are not ruled by the Arnold-Liouville theorem (see \cite{MZ}), since the level set of $\mathbb{F}:=(F_1,\dots,F_n)$ are not necessarilly of dimension $n$ .
In the following, we investigate briefly several dynamics of the last type, corresponding to different diophantine assumptions on the $w_i$'s. We illustrate the preceeding remark showing that it is not necessary that periodic orbits arise in $n$-dimensional tori, and that they actually provide us with a various zoology of dynamics concerning families of periodic orbits.\

But before we recall some straightforward computations to solve (\ref{hamsyst}) very explicitly in our case.
Here, the dynamical system is linear and we have
$$M_z(t)=:M(t)=\left(
\begin{array}{c|c}
\cos(t\sqrt{S})           & \sqrt{S}^{-1}\sin(t\sqrt{S})\\\hline
 -\sqrt{S}\sin(t\sqrt{S}) & \cos(t\sqrt{S})              \\
\end{array}
\right).$$
So the flow is given by $\Phi_t(z)=M(t)z$, and the solutions of (\ref{hamsyst}) by:
\beq
\left\{
\barr{l}
x_j(t)=\cos(w_jt) x_j(0)+ w_j^{-1}\sin(w_j t) \xi_j(0).\\
\xi_j(t)=-w_j\sin(w_j t) x_j(0) + \cos(w_j t) \xi_j(0).
\earr\right.
\eeq
Note that the set of periods of the dynamics is:
\beq
\mathcal{P}:=\overset{d}{\underset{j=1}{\bigcup}} \frac{2\pi}{w_j}\zz.
\eeq
So if $T\in \mathcal{P}$, then the set of all $z$ for which $T$ is a period is given by
\beq\label{deltaT}
\Delta_T:=\underset{j:w_j T\in 2\pi \zz}{\bigoplus} \rr e_j+\rr e_j^{\prime},
\eeq
where $(e_1,\dots,e_n;e_1^{\prime},\dots,e_n^{\prime})$ is the canonical basis of $\rdd$. Note that, if $z\in\Delta_T$, then $\grad\in\Delta_T$. Thus $\Delta_T$ and $\nrj$ are transverse submanifolds in $\rdd$. We can already notice that if $q\in\{1,\dots,n\}$, if we choose $w_1=\dots=w_q=1$ and all others $w_j$ out of $\nn$, then $\dim(\Delta_{2\pi}\cap\nrj)=2q-1$. This means that, within this family of quadratic Hamiltonians, we can find some  $\ce$ with isolated connected components of any dimension between $1$ and $2n-1$ . Note also that after a short calculation , if $z\in \Delta_T$, we have $E_1(T,z)=\ker(M(T)-Id)$, and thus (\ref{hypDG}) is satisfied.
\subsubsection{Different types of flows near periodic orbits}
The proofs of the following lemmata are left to the reader.
We start with the case owning the smallest number of periodic orbits.
\begin{lem}
The following assertions are equivalent:\\
(1) $\forall i\neq j, $ $\frac{w_i}{w_j}\notin \mathbb{Q}$.\\
(2) All periodic orbits of $H_w$ are non-degenerate.
\end{lem}
\noindent{\it Example}: $n=2$, $w_1=1$, $w_2=\sqrt{2}$.\\

Then we analyse the opposite case:
\begin{lem}
The following assertions are equivalent:\\
(1) $\forall i, j, $ $\frac{w_i}{w_j}\in \mathbb{Q}$.\\
(2) All orbits of $H_w$ are periodic.\\
(3) $\exists T\in \mathcal{P}\setminus\{ 0 \}$ s.t. $\forall j, w_j T\in 2\pi \mathbb{Z}$.\\
(4) $\exists z\in\rdd$ s.t. $rank(\nabla F_1(z),\dots,\nabla F_n(z))=n$, with $z$ periodic point.\\
(5) $\exists E> 0$ s. t. the flow is periodic on $\nrj$.
\end{lem}
\noindent{\it Example}: $n=2$, $w_1=1$, $w_2=2$,
Note that here, periodic orbits can have different primitive periods. In our exemple, $T_1=2\pi$, $T_2=\pi$ are two different primitive periods.
In this case we could have applied the Arnold-Liouville theorem.\\

The very extrem is the case of the harmonic oscillator:
\begin{lem}
The following assertions are equivalent:\\
(1) $w_1=\dots=w_n$.\\
(2) All orbits of $H_w$ are periodic with the same period.\\
(3) $\exists E> 0$ s. t. on $\nrj$ all orbits of $H_w$ are periodic with the same period..
\end{lem}
Some cases in between are given in terms of symmetries, where periodic orbits should arise in manifolds of various dimensions. For a general Hamiltonian $H$ we define the maximal group  $G_{max}$ of symplectic symmetries by
$$G=G_{max}:=\{ g\in Sp\, (n) \; s.t. \; H(gz)=H(z), \;\; \forall z\in \rdd \}.$$
where $Sp\, (n)$ is the group of symplectic matrices of size $2n$. $G_{max}$ is obviously a subgroup of $Sp\, (n)$. Let us suppose that it is a compact Lie group. The invariance of $H$ by the group classically implies that the flow $\Phi_t$ of $H$ commutes with all elements of $G_{max}$ (these considerations are more detailled in section \ref{compactsymmetries}). As a consequence, if $(T,z)$ is a periodic point, then so is $(T,gz)$. Thus, the whole manifold $G(\gamma_z)$ is made of periodic orbits of same period and is included in $\nrj$ where $E:=H(z)$. We now introduce a rough notion which ensures that these "tubes" of periodic orbits are "isolated" in $\rr\times \nrj$. We should enlarge this notion in section \ref{conservative}. If $(T,z)$ is a periodic orbit of $H$, we say that $(T,z)$ is "{\it NDR in its energy surface}" if we have:
$$\dim E_1(T,z)=\dim(\rr \jgrad + \mathcal{G}(z)) +1.$$
where $\mathcal{G}$ is the Lie algebra of $G$.  This boils down to assuming that a periodic point in $\nrj$ becomes either a non-degenerate periodic orbit or a equilibrium point in $\nrj/G$.  When $G_{max}=\{Id\}$, we find back the notion of non-degenerate periodic orbit. The notation NDR stands for "Non-Degenerate in the Reduced space $\nrj/G$".
In the case of the quadratic Hamiltonian $H_w$,
$$G_{max}=Sp\,(n)\cap [(\sqrt{A_0})^{-1}O(2n)\sqrt{A_0}],\mbox{ where }A_0=\left(
\barr{cc}
S & 0\\
0 & I_d
\earr\right).$$
One can easily show that all periodic orbits are actually equilibrium points in the quotient. Thus, periodic orbit will appear by isolated `tubes' of the form $G(z)$ in $\nrj$.
\begin{lem}\label{NDRquadrat}
The following assertions are equivalent:\\
(1) $\forall i,j$, $\frac{w_i}{w_j}\in \mathbb{Q}\Rightarrow w_i=w_j$.\\
(2) All periodic orbits of $H_w$ are NDR in their energy surface.\\
\end{lem}
\noindent{\it Example}: $d=3$, $w_1=w_2=1$, $w_3=\sqrt{2}$.
\subsubsection{STF for quadratic Hamiltonians}
We use the notations of section \ref{STFtheo}.
\begin{prop}\label{STFquadrat}
For any $w\in(\rr_+^*)^d$, and any $E$ in $\rr^*$, we have when $h\to 0^+$
\beq
\mathcal{G}_E(h)=\sum_{T\in \overset{d}{\underset{j=1}{\bigcup}} \frac{2\pi}{w_j}\zz}(2\pi h)^{1-R(T)}\frac{e^{\frac{iTE}{h}}\hat{f}(T)\psi(E)e^{i\frac{\pi}{2}\sigma_T}}{2\pi\displaystyle{ \prod_{Tw_j\notin 2\pi\zz} \sqrt{2}|1-\cos(Tw_j)|^{1/2} }}
\int_{\Delta_T\cap\nrj} \frac{d\sigma_{\Delta_T\cap\nrj}}{\norm{\nabla H(z)}}+O(h^{2-R(T)}).
\eeq
where  $\Delta_T$ is given by (\ref{deltaT}), $R(T):=card \{ j:w_j T\in 2\pi \zz\}$, $d\sigma_{\Delta_T\cap\nrj}$ is the Riemannian measure on $\Delta_T\cap\nrj$, and $\sigma_T$ is the common Maslov index of trajectories of $\Delta_T\cap\nrj$.
\end{prop}
\begin{proof}
$\mathcal{L}_E$ is finite since $\mathcal{P}$ is discrete. Moreover, for all $T$ in $\mathcal{L}_E$, we have seen that
$$\mathcal{Z}_T=\Delta_T\cap\nrj, \quad \forall z\in \mathcal{Z}_T,\quad E_1(z,T)=\Delta_T=\ker(M_z(T)-I).$$
Besides, $\nrj$ and $\Delta_T$ being transversal manifolds in $\rdd$, $\Z_T$ is a submanifold of $\rdd$ and
$\forall z\in\mathcal{Z}_T$, $T_z \mathcal{Z}_T=\ker(M_z(T)-I)\cap T_z\nrj$. We clearly have (\ref{hypDG}). Hence, we can apply (\ref{DGsimple}).
An orthonormal basis of $E_1$ is given by $\beta=\vect\{e_j,e_j':Tw_j\in 2\pi \zz\}$. Since $w_0(e_i',e_j)=\delta_{i,j}$, we obtain $\det({w_0}_{|_{E_1}})=1$. Performing a computation of eigenvalues of $M_z(T)$, we obtain $(e^{\pm iTw_j})_{\{j=1,\dots,n\}}$. We get
$$d(T,z)^{2}=\frac{(-1)^{n+R(T)}}{\norm{\nabla H(z)}^2\underset{Tw_j\notin 2\pi\zz}{\prod} 2[1-\cos(Tw_j)] }.$$
Endly, the computation of $\A(T,Y)$ is also achieved using the expression of the flow.
\end{proof}
\section{Conservative systems}\label{conservative}
In this section we give practical means to obtain systems satisfying hypotheses of Theorem \ref{gutztheo2} for conservative systems, i.e. systems with several independent constants of motion near $\nrj$.
\subsection{Normal periodic orbits}
Let us note that if $F:\rdd\to\rr$ is a first integral of the system (\ref{hamsyst}), and if $(T_0,z_0)$ is a periodic orbit of $\nrj$, then, differentiating 
$$F(\Phi_{T_0}(z))=F(z),$$
with respect to variable $z$ at $z_0$, we obtain that $\nabla F(z_0)\in\ker( ^tM_{z_0}(T_0)-Id)$, and since $M_{z_0}(T_0)$ is symplectic, this means that
\beq
J\nabla F(z_0)\in\ker(M_{z_0}(T_0)-Id).
\eeq
This makes $\ker(M_{z_0}(T_0)-Id)$ bigger for conservative systems. Moreover, $E_1$ also grows drastically in this case as shows Lemma \ref{Meyer}, above all when first integrals are in involution.
If our first integral $F$ is independent of $H$ at $z_0$, applying this lemma to $M=M_{z_0}(T_0)$, $V= \vect (J\nabla F(z_0),J\nabla H(z_0))$, $W=\rr J\nabla H(z_0)$, we obtain that $\dim E_1\geq 3$. Thus, for conservative systems, we won't have non-degenerate periodic orbits.
Another way to see this is to note that, if $\Phi^F_t$ denotes the flow of $F$, then since it commutes with the one of $H$, all points of the image of $t\mapsto \Phi^F_t(z_0)$ are $T_0$-periodic of $\nrj$. This is in contradiction with the fact that non-degenerate periodic orbit $\nrj$ are isolated in $\rr\times\nrj$.\

Our aim in this section is to broaden the notion of non-degenerate periodic orbit in order to be able to apply it to conservative systems. This notion should also ensure that the flow is clean on $\nrj\times\supp \hat{f}$. For this, we make use of the concept of normal periodic orbits introduced in \cite{Vdb} and \cite{SMK}. Let $\Sigma$ be a smooth manifold, and $X$ a vector field on $\Sigma$ with flow $\Phi_t$. Let $(T,x)$ be a periodic orbit of $X$. Suppose that there exist first integrals $F_1,\dots,F_k$ in a neighbourhood of $\gamma_x$ in $\Sigma$. Then, by using the same argument as in the beginning of this section, we always have
$$(M_x(T)-Id)(T_x\Sigma) + \rr X(x)\subset \bigcap_{j=1}^k \ker d_x F_j.$$
\begin{dfn}
We say that $(T,x)$ is normal if there exist first integrals $F_1,\dots,F_k$ in a neighbourhood of $\gamma_x$ in $\Sigma$ such that $d_xF_1,\dots,d_x F_k$ are independent and that
$$(M_x(T)-Id)(T_x\Sigma) + \rr X(x)=\bigcap_{j=1}^k \ker d_x F_j.$$
where $M_x(T):=\partial_x \Phi_T (x)$.
\end{dfn}
A slight extension of the paper \cite{Vdb} to manifolds leads to the following
\begin{theo}\cite{Vdb}\label{Vdbth}
Using the same notations, let $(T,x)$ be a normal periodic orbit for $X$ on $\Sigma$. Then there exists a neighbourhood $U$ of $\gamma_x$ in $\Sigma$, $\e>0$, and smooth submanifold  $\mathcal{P}$ of $]T-\e,T+\e[\times U$ of dimension $k+1$, such that $(T',x')\in ]T-\e,T+\e[\times U \mbox{ and } \Phi_{T'}(x')=x'$ if and only if $(T',x')\in \mathcal{P}$.
\end{theo}
This is a kind of generalization of the `cylinder theorem' (see \cite{MH} or \cite{MZ}) to normal periodic orbits. Note that in the case of the system (\ref{hamsyst}), using symplecticity of the linearized flow, $(T,z)$ is  normal periodic orbit in $\rdd$ if and only if there exists first integrals $F_1,\dots,F_k$ on a neighbourhood of $\gamma_z$ such that $\nabla F_1(z),\dots,\nabla F_k(z)$ are independent and that
\beq\label{eqnormaleucl}
\ker(M_z(T)-Id)\cap T_z\nrj =\vect(J\nabla F_1(z),\dots,J\nabla F_k(z)),
\eeq
and in general we always have
$$ \vect(J\nabla F_1(z),\dots,J\nabla F_k(z)) \subset \ker(M_z(T)-Id)\cap T_z\nrj.$$
We introduce a new definition close to the preceeding notion
\begin{dfn}\label{sigmanormal}
We say that $(T,z)\in \rr\times\nrj$ is "$\nrj$-normal" if there exists  first integrals $F_1,\dots,F_k$ on a neighbourhood of $\gamma_z$ in $\rdd$ such that $\nabla F_1(z),\dots,\nabla F_k(z)$ are independent and that
$$(M_z(T)-Id)(T_z\nrj) + \rr \jgrad= \vect(\nabla F_1(z),\dots,\nabla F_k(z))^{\perp_{\rdd}}.$$
\end{dfn}
Note that this notion is slightly different from saying that $(T,z)$ is normal for the restriction of (\ref{hamsyst}) to $\nrj$ since we suppose a little bit more assuming that our first integrals are defined not only on $\nrj$ but on an open set of $\rdd$. As a  consequence, periodic orbits of the associated $\mathcal{P}$ in $\nrj$ will have the same period $T$ (see Proposition \ref{normalFP}).
\begin{lem}\label{critere}
$(T,z)\in \ce$ is $\nrj$-normal if and only if $(T,z)$ is normal for the system (\ref{hamsyst}) in $\rdd$ and $\jgrad \notin (M_z(T)-Id)(T_z\nrj)$.
\end{lem}
\begin{proof}
If $(T,z)\in \rr\times\nrj$ is $\nrj$-normal, then it is of course normal for the system (\ref{hamsyst}) in $\rdd$. Reasoning by contradiction, suppose that $\jgrad \in (M_z(T)-Id)(T_z\nrj)$. Let us denote by $\tm$ the restriction of $M_z(T)$ to $T_z\nrj$. Then $\mbox{rank}(\tm-Id)=2n-k$, and by the rank theorem, $\dim(\ker(\tm-Id))=k-1$. This is in contradiction with the fact that $\ker(\tm-Id)$ already contains the $k$-dimensional vector space generated by the vectors $J\nabla F_1(z),\dots,J\nabla F_k(z)$.\

Reciprocally, suppose that $(T,z)$ is normal for the system (\ref{hamsyst}) in $\rdd$ and that $\jgrad \notin (M_z(T)-Id)(T_z\nrj)$. Let $W(z):=\vect(\nabla F_1(z),\dots,\nabla F_k(z))$. We have 
$\ker(M_z(T)-Id)\cap T_z\nrj=JW(z)$. By the rank theorem, we get $\dim (M_z(T)-Id)(T_z\nrj)=2n-k-1$, thus $\dim (M_z(T)-Id)(T_z\nrj)+\jgrad=2n-k$. But since
$$(M_z(T)-Id)(T_z\nrj) + \rr \jgrad\subset (M_z(T)-Id)\rdd + \rr \jgrad=W(z)^{\perp},$$
we have equality for dimensional reasons.
\end{proof}
\begin{prop}\label{normalFP}
Suppose that periodic orbits of $\supp\hat{f}\times\nrj$ are all $\nrj$-normal. Then $\mathcal{L}_E$ is finite and for all $T\in\mathcal{L}_E$, $\mathcal{Z}_T$ has a finite number of connected components. These connected components are submanifolds of $\nrj$. Moreover the flow is clean on $\nrj\times\supp\hat{f}$,
and for all $T\in\mathcal{L}_E$, for all $z\in\mathcal{Z}_T$, $\jgrad \notin (M_z(T)-Id)(T_z\nrj)$.
\end{prop}
\begin{proof}
We first prove a local lemma
\begin{lem}\label{localnormalFP}
If $(T,z)\in\ce$ and if $F_1,\dots,F_k$ are the $k$ independent first integrals of Definition \ref{sigmanormal}, then $\exists \e>0$, $\exists U_E$ neighbourhood of $\gamma_z$ in $\nrj$, $\exists \Gamma_E$ submanilfold of $U_E$ such that 
$$\ce\cap ]T-\e,T+\e[\times U_E=\{T\}\times \Gamma_E.$$
Moreover $\dim\Gamma_E=k$ and $\forall z'\in\Gamma_E$, $T_{z'}\Gamma_E=J\vect(\nabla F_1(z'),\dots,\nabla F_k(z'))$.
\end{lem}
\begin{proof}
We can apply Theorem \ref{Vdbth} to the restriction of (\ref{hamsyst}) to $\nrj$. Of course $H$ doesn't count as a first integral. We obtain a neighbourhood $U$ of $z$ in $\nrj$ and $\e>0$ such that $\ce\cap (]T-\e,T+\e[\times U)$ is a manifold of dimension $k$. If $i\in\{1,\dots,k\}$, we denote by $\Phi^{F_i}_t$ the flow of the Hamiltonian dynamical system generated by $F_i$. Note that since $t\mapsto (T,\Phi^{F_i}_t(z))$ is a smooth path in $\ce\cap (]T-\e,T+\e[)\times U)$, we have $T_{(T,z)} \ce=\{0\}\times JW(z)$ where $W(z):=\vect(\nabla F_1(z),\dots,\nabla F_k(z))$. Since the gradients of $F_1,\dots,F_k$ are still linearly independent close to $z$, for a smaller $U$, we can suppose that we have
$$\forall (T',z')\in\ce\cap(]T-\e,T+\e[)\times U),\quad T_{(z',T')}\ce=\{0\}\times JW(z').$$
Using a chart of $\ce\cap (]T-\e,T+\e[)\times U)$ defined on a connected domain, it is then easy to show that, schrinking $]T-\e,T+\e[)\times U$,  we have that $\ce\cap (U\times ]T-\e,T+\e[)$ is of the form  $\{T\}\times\Gamma_E$.
\end{proof}
Then, using the compacity of $\supp\hat{f}\times\nrj$, we obtain easily that $\mathcal{L}_E$ is finite (reason by contradiction).
The set $\mathcal{Z}_T$ satisfies the property
$\forall x\in \Z_T,$ $\exists k_x\in\nn,$ $\exists U_x$  open set of $\nrj$ such that $\Z_T\cap U_x$ is a submanifold of dimension $k_x$ of $\nrj$.
From this it follows easily that each connected component of $\Z_T$ is a submanifold of $\nrj$. Now to show that $\Z_T$ has only a finite number of connected components, let us suppose by contradiction that we have a sequence $(C_n)_{n\in\nn}$ of disjoint connected components of $\Z_T$. Pick a $z_n$ on each $C_n$. Then by compacity of $\nrj$ and thus of $\Z_T$, by extracting a subsequence, one can always suppose that $z_n$ converges to $z^*\in\Z_T$. Applying the preceeding property to $z^*$, one can always suppose that $\Z_T\cap U_{z^*}$ is connected. Thus for $n$ great enough, $z_n\in U_{z^*}$, and $C_n$ is the connected component of $\Z_T$ containing $z^*$. This contradicts the fact that $C_n$ are all disjoints.
Endly, Lemma \ref{localnormalFP} garanties us that $\forall (T,z)\in \ce$, we have $\E_1(T,z)=T_z\Gamma_E$. Thus the flow is clean on $\nrj\times\supp\hat{f}$.
\end{proof}
We see that under normal periodic orbits assumptions, the hypotheses of the first part of Theorem \ref{gutztheo2} are fulfilled. Note that a non-degenerate periodic orbit $(z,T)$ is $\nrj$-normal. We have even a more general statement:
\begin{lem}
Let $(T,z)$ be  periodic orbit of $\nrj$ such that there exist independent first integrals $F_1,\dots,F_k$ is a neighbourhood of $\gamma_z$ in $\rdd$ and such that $\dim E_1(T,z)=k+1$. Then $(T,z)$ is a $\nrj$-normal periodic orbit.
\end{lem}
An example of such a situation is given by Lemma \ref{NDRquadrat}.
\begin{proof}
First note that, without any normality assumption, $E_1$ is never included in $\nrj$. Indeed, $E_1\subset\nrj\iff \jgrad\in (JE_1)^\perp=V_1$, and we have seen that $E_1\cap V_1=\{0\}$. Now, under our assumptions, we have thus $\dim (E_1\cap\nrj)=k$ and
$$\vect(J\nabla F_1(z),\dots,J\nabla F_k(z)) \subset \ker(M_z(T)-Id)\cap T_z\nrj \subset E_1 \cap T_z\nrj.$$
For dimensional reasons, we have equality everywhere and thus $(T,z)$ is normal in $\rdd$. Moreover, if $\alpha\in T_z\nrj$ is such that $\jgrad=(M_z(T)-Id)\alpha$, then $\alpha\in E_1\cap \nrj\subset \ker(M_z(T)-Id)\cap T_z\nrj$, thus $\jgrad=0$, which is excluded. The lemma is proved using Lemma \ref{critere}.
\end{proof}
Note that, even if the assumption `$\dim E=k+1$' seems a more straightforward generalization of the concept of non-degenerate periodic orbit, in view of Lemma \ref{Meyer}, it will actually not be fulfilled each time that the system has first integrals different of $H$ in involution, i.e. satisfying $\{F_i,F_j\}=0$, where $\{,\}$ is the Poisson bracket.\

To end this subsection, we give an example of a periodic orbit which in {\it not} normal. Let us consider the quadratic Hamiltonian of section \ref{classicalexamples} in dimension $n=2$, with $w_1=1$, and $w_2=2$. In this case, we have two periods, $\pi$ and $2\pi$, for which $\Delta_\pi=\rr e_2+\rr e_2'$ and $\Delta_{2\pi}=\rr^4$. If we consider the period $T=2\pi$ with $z\in\Delta_\pi$, we have $\Phi_T(z)=z$, and, for any first integral $F$ around $\gamma_z$, we have $J\nabla F(z)\in\Delta_\pi=\rr e_2+\rr e_2'$. But we have $\E_1(T,z)= T_z\nrj$. Thus (\ref{eqnormaleucl}) cannot be fulfilled for any system of first integrals.
\subsection{Compact symmetries}\label{compactsymmetries}
A nice way to obtain first integrals of Hamiltonian dynamical systems is due to E. Noether: one can look for the symmetries of the Hamiltonian $H:\rdd\to\rr$. We suppose that there is a compact Lie group $G\subset Sp\, (n)$\footnote{$Sp\, (n)$ is the group of symplectic linear maps of $\rdd$.} such that 
\beq\label{symgroup}
\forall g\in G, \quad  H(gz)=H(z).
\eeq
Actually, we should be concerned by groups of symplectic diffeomorphisms and not only linear groups. We do this for sake of simplicity. Note that since a group of isometries is affine, with our compactness assumption, we are not far from reaching all symmetries of the problem. Moreover, for linear groups, Noether first integrals have an explicit expression: if $A\in \mathcal{G}$ the Lie algegra of $G$, then
\beq
F_A(z):=<JAz,z>_{\rdd},
\eeq
is a first integral of the system (\ref{hamsyst}). Indeed, by derivating with respect to variable $t\in\rr$ at $t=0$ the identity $H(e^{tA}z)=H(z)$, we obtain $\{H,F_A\}=0$ since $JA$ is symmetric. Now, if $g\in G$, differentiating the identity (\ref{symgroup}) with respect to variable $z$, we obtain, using the symplecticity of $G$ that
$$\forall z\in \rdd,\quad g\jgrad=J\nabla H(gz).$$
As a consequence, the flow of $H$ commutes with all elements of $G$. Now, if $(T,z)\in\rr\times\nrj$ is a periodic orbit, then for all $g$ in $G$, so is $(T,gz)$. We obtain a manifold $\{T\}\times G(\gamma_z)$ included in $\ce$. Note that when our group is not finite, this proscribes non-degenerate periodic orbits. We have in mind to ensure that $\{T\}\times G(\gamma_z)$ is a connected component of $\ce$ and that we get a clean flow at $(T,z)$. In view of definition \ref{def1}, a necessary condition for that is: 
\beq\label{normalgroup}
\left\{
\begin{array}{l}
\ker(M_z(T)-Id)\cap T_z\nrj=\rr\jgrad +\alg z.\\
\jgrad\notin (M_z(T)-Id)T_z\nrj.
\end{array}\right.
\eeq
Morally speaking, this means that $G$ has to be big enough so that $\rr\jgrad +\alg z$ would `fill' $\ker(M_z(T)-Id)\cap T_z\nrj$, or from an other point of view, that $\ker(M_z(T)-Id)$ must not be too big. This condition is actually enough to ensure a clean flow as states the following
\begin{prop}\label{STFgroup}
Suppose that $H$ admits a compact connected Lie group of symmetry $G$ such that $\ce$ satisfies (\ref{normalgroup}). Then there exist periodic points $z_1,\dots,z_r$ of $\nrj$ such that 
\beq\label{cecompact}
\ce=\bigcup_{i=1}^r\bigcup_{\substack{ n\in\zz: \\ nT_{z_i}^*\in\mbox{{\scriptsize supp}}\hat{f}} }
 \{nT_{\gamma_{z_i}}^*\} \times G(\gamma_{z_i}),
\eeq
where $T_{z_i}^*$ is the primitive period of $z_i$. Moreover the flow is clean on $\supp\hat{f}\times \nrj$, and if (\ref{hypRC}) is fulfilled on $\supp\hat{f}\times \nrj$, then, assuming for simplicity that $0\notin \supp\hat{f}$, one has when $h\to 0^+$
$$\mathcal{G}_E(h)=\psi(E)\sum_{i=1}^r\sum_{\substack{ n\in\zz: \\ nT_{z_i}^*\in\mbox{{\scriptsize supp}}\hat{f}} } (2\pi h)^{\frac{1-k_i}{2}} \frac{e^{\frac{i}{h}n\A_i}\hat{f}(nT_{z_i}^*)e^{i\frac{\pi}{2}\sigma_{i,n}}}{\sqrt{|\det(M_{z_i}(T_{z_i}^*)^n-Id)_{|_{V_1}}|}} $$
\beq\label{STFgroup2}
\times\int_{G(\gamma_{z_i})} \frac{|\det({w_0}_{|_{E_1}})|^{\ud}}{\norm{\Pi_{E_1}(\grad)}}\frac{d\sigma_{G(\gamma_{z_i})(z)}}{\sqrt{|\det(\Pi_{\E_5}J(M_z(T)-Id)_{|_{\E_5}})|}} 
+O(h^{\frac{1-k_i}{2}+1}),
\eeq
where $k_i=\dim G(\gamma_{z_i})$, $\A_i:=\int_{\gamma_{z_i}}pdq$, $d\sigma_{G(\gamma_{z_i})(z)}$ is the Riemannian measure on $G(\gamma_{z_i})$, $\sigma_{i,n}\in\zz$, and other terms are defined in Theorem \ref{gutztheo}.
\end{prop}
\begin{proof}
From Proposition \ref{normalFP} and Lemma \ref{localnormalFP}, we get that $\mathcal{L}_E$ is finite, the flow is clean and for all $(T,z)$ in $\ce$, $\dim(\Z_T)=\dim(G(\gamma_z))$. As $G(\gamma_z)$ is a compact submanifold of same dimension as $\Z_T$, it is one of is connected components. This yields (\ref{cecompact}). For the rest of the theorem, one applies Theorem \ref{gutztheo2}, noting that if $n\in \zz$, if $z\in G(\gamma_{z_i})$, then the matrices  $M_z(nT_{z_i})$ and $M_{z_i}(T_{z_i})^n$ are conjugated. Indeed, if $g\in G$ and $t\in\rr$, we have
$$M_{gz}(T)=gM_z(T)\gmu,\quad M_{z}(T)^n=M_z(nT),\quad M_{\Phi_t(z)}(T)=M_z(t)M_z(T)M_z(t)^{-1}.$$
This is obtained differentiating with respect to the $z$ variable the identities $\Phi_{t+s}(z)=\Phi_s(\Phi_t(z))$ and $g\Phi_t(z)=\Phi_t(gz)$.
\end{proof}
Such a situation is of course fulfilled in Lemma \ref{NDRquadrat}. An example where it is not fulfilled is the same as in the previous subsection.
If we compare formula (\ref{STFgroup2}) to the one of the non-degenerate case (see (\ref{gutzND})), we see that we obtain similar quantities, unless for the term $T_\gamma^*$, which is here replaced by our integral on $G(\gamma_{z_i})$. Of course, the ideal result would be to interpret this integral in terms of geometrical quantities describing $G(\gamma_{z_i})$. A heuristical interpretation can be found in the work of physicists Creagh and Littlejohn (\cite{CL1},\cite{CL2}). So far we have not been able to achieve this in a rigorous general mathematical framework for the moment. Nevertheless, we do so in the next section for a class of integrable systems.
\section{Integrable case: the formula of Berry-Tabor}\label{berrytabor}
In this section we give a trace formula for a certain class of integrable systems in terms of action-angle variables. In this context, our result gives actually less informations on the spectrum than the Bohr-Sommerfeld conditions (see for example \cite{CdV3}\cite{Do2}\cite{AC}) which allow to localize the spectrum, whereas we are dealing with spectral statistics. Nevertheless, our goal is to highlight the geometrical interpretation of the STF, in order to be able to do so in more general contexts, where less is known about the spectrum. Another paper by Charbonnel-Popov \cite{CP} may seem very close to what we are presenting here. However, the object of study is not the same since they are considering trace formulae defined with Schwartz functions in $\rr^k$ of $k$ commuting operators, which can not be reduced to a Schwartz function of only one of them. We should here recover results of physicists Berry and Tabor (\cite{BT1}\cite{BT2}) in a rigorous way. To our knowledge, these results seems to be strangely absent from the mathematical literarure.
\subsection{Assumptions and notations on the integrable system}\label{intblabla}
Let $H:\rdd\to\rr$ be a smooth Hamiltonian, $E\in \rr$. Suppose that there exists $\e_0>0$ such that $H^{-1}([E-\e_0,E+\e_0])$ is compact without critical points of $H$ and that the system is Liouville integrable (or integrable) on $U_E:=H^{-1}(]E-\e_0,E+\e_0[)$, i.e. assume that there exist $n$ smooth constants of motion, $F_j:\rdd\to\rr$,  $j=1,\dots,n$, such that we have:
\beq\label{integrable}
\left\{
\barr{l}
(1) \qquad \{F_i,F_j\}=0, \;\; \forall i,j=1,\dots,n, \mbox{ on } U_E.\\
(2) \qquad \nabla F_1(z),\dots,\nabla F_n(z) \mbox{ are independent for all $z$ in $U_E$.}
\earr\right.
\eeq
where $\{,\}$ is the Poisson bracket. We can always assume that $F_1=H$. We denote the flow of $F_j$ by $\Phi_t^{F_j}$. These different flows commute since the integrals are in involution. Let
$$\mathbb{F}:=(F_1,\dots,F_n):U_E\to \rd.$$
and if $c\in\rd$, $\mathcal{T}_c:=\mathbb{F}^{-1}(\{c\}).$
Due to the independence of the first integrals, each $\mathcal{T}_c$ is a compact submanifold of dimension $n$ of $\rdd$. We will suppose that
\beq\label{connectedness}
\mbox{For all $c$ in $\rd$, $\mathcal{T}_c$ is connected.}
\eeq
According to Arnold \cite{A}, each $\mathcal{T}_c$ is diffeomorphic to a torus and it is covered by the joint flow of the  $F_j$. We obtain also local action-angle coordinates near each $\mathcal{T}_c$ (see \cite{MZ}). Nevertheless, we will suppose that we have {\it global} action-angle coordinates on $U_E$. To this purpose we make the following hypothesis
\beq\label{globalAA}
\mathbb{F}(U_E) \mbox{ is simply connected in $\rd$.}
\eeq
According to Duistermaat \cite{D}, we have global action-angle coordinates on $U_E$, i.e.
there exists a {\it symplectic} diffeomorphism $\psi:U_E\to \mathbb{T}^n\times D$, where $\mathbb{T}^n:=\rr^n/(2\pi \zz)^n$, $D$ is an open set in $\rd$ such that, if we write $\psi=(\psi_1,\psi_2)$
\begin{itemize}
\item $\forall z,z'\in U_E$, $\;\mathbb{F}(z)= \mathbb{F}(z') \iff \psi_2(z)=\psi_2(z')$.
\item If $\vfi_t=(\theta_t,I_t):=\psi\circ \Phi_t\circ \psi^{-1}$, then for all $(\theta,I)\in \mathbb{T}^n\times D$ we have
\beq\label{SDint}
\left\{
\begin{array}{l}
\dot{\theta}_t(\theta,I)=\nabla \hta(I).\\
\dot{I}_t(\theta,I)=0.
\end{array}\right.
\eeq
\end{itemize}
where $\hta:D\to\rr$ is the smooth function defined by
$$\hta(I):=H\circ\psi^{-1}(\theta_0,I),$$
whatever $\theta_0$ in $\mathbb{T}^n$, since $H$ is constant on the tori. Thus $\psi$ transform symplectically the motion into translations on tori. In other words, the flow $\vfi_t$ is given by:
\beq\label{flotint}
\vfi_t(\theta,I)=(\theta+[t\nabla \hta(I)],I)
\eeq
where $[.]:\rd\to \mathbb{T}^n$ is the canonical projection.
One usually denotes by $w:D\to \rd$ the frequency function defined by
\beq\label{frequence}
w(I):=\nabla \hta(I).
\eeq
 We suppose that our integrable system is {\it non-degenerate}, i.e.
\beq\label{NDint}
\forall I\in D, \; \det(w'(I)))\neq 0.
\eeq

This assumption is `generically' satisfied and it is also often used for perturbation of integrable systems (see \cite{A}).
Let us now focus on periodic orbits of the system. Of course if one point of a torus is periodic, then all are. We will talk about `periodic tori'. From (\ref{flotint}) we deduce that for $z$ in $U_E$, $T\in\rr$, we have $\Phi_T(z)=z$ if and only if $T w(I)\in (2\pi\zz)^n$, where $\psi_2(z)=:I$. Using this remark and assumption (\ref{NDint}), one can apply implicit function theorem to the function $f:\rr\times D\to \rd$ defined by $f(T,I)=Tw(I)$, to prove that, if $(T_0,I_0)$ is a periodic torus, then the set of periodic tori around $(T_0,I_0)$ in $\rr\times D$ is locally given by the graph of a function $\mathcal{I}:]T_0-\e,T_0+\e[\to D$ such that 
\beq\label{tfi}
\forall T\in ]T_0-\e,T_0+\e[,\quad Tw(\mathcal{I}(T))=T_0 w(I_0).
\eeq
We introduce
\beq\label{nrja}
\nrja:=\{\hta=E\}\subset \rd.
\eeq
Note that since our dynamical system has no equilibrium point on $\nrj$, we have that $\forall I\in \nrja$, $\nabla\hta(I)\neq 0$, and thus $\nrja$ is a hypersurface of $\rd$.
To enter the context of Theorem \ref{gutztheo2}, we need to have a constant period on $\nrja$. Thus we would like the image of $\mathcal{I}$ to be transversal to $\nrja$ at $I_0$ in $\rd$, for which it is enough to have $<\mathcal{I}'(I_0),\nabla \hta(I_0)>\neq 0$. But derivating equation (\ref{tfi}), we obtain that $\mathcal{I}'(T_0)=-\frac{1}{T_0}w'(I_0)^{-1}(w(I_0))$. This leads us to make the following hypothesis
\beq\label{isoint}
\forall (T,I)\in\supp \hat{f}\times \nrja \mbox{ such that } Tw(I)\in (2\pi\zz)^n, \quad <w'(I)^{-1}(w(I)),w(I)>\neq 0.
\eeq
We have proved the following
\begin{lem}\label{toresisoles}
Make assumptions (\ref{NDint}) and (\ref{isoint}). Let $(T_0,z_0)\in\supp \hat{f}\times\nrj$ such that $\Phi_{T_0}(z_0)=z_0$. Then it exists $\e>0$ such that if $(T,I)\in \supp\hat{f}\times\nrja$,
if $|T-T_0|+|I-I_0|\leq \e$ and if $(T,I)$ is a periodic torus then $I=I_0$ and $T=T_0$. As a corollary, there exists only a finite number of periodic tori of $\nrj$ with period in $\supp\hat{f}$.
\end{lem}
The STF formula should involve the function $K:\nrja\to\rr$ of {\it Gaussian curvuture} of $\nrja$. We recall (see \cite{spivak} p.3B-21 5.Corollary) that if $I\in \nrja$, if $\vfi$ is a parametrization of $\nrja$ defined on a neighbourhood of zero in $\rr^{n-1}$ such that $\vfi(0)=I$, then $K(I)$ is defined by
\beq\label{curv}
K(I):=\frac{ \det\left( <\frac{\nabla \hta(I)}{|\nabla\hta(I)|},\frac{\partial^2\vfi}{\partial_{\xi_i}\partial_{\xi_j} }(0)> \right)_{1\leq i,j\leq n-1} }{ \det\left(<\partial_{\xi_i}\vfi(0),\partial_{\xi_j}\vfi(0)>\right)_{1\leq i,j\leq n-1}}.
\eeq
The following lemma shows that hypotheses (\ref{NDint}) and (\ref{isoint}) are actually mandatory if we want the curvature to be non-zero.
\begin{lem}\label{courbur}
For all $I\in\nrja$, 
\beq\label{courbformule}
\det(w'(I))<w'(I)^{-1}(w(I)),w(I)>=|w(I)|^2K(I)(-1)^{n-1}|w(I)|^{n-1}.
\eeq
\end{lem}
\begin{proof}
Let $\vfi$ be a chart of $\nrja$ at $I$ as in the definition (\ref{curv}). Differentiating at $\xi=0$ the relation
\beq\label{eqorth}
<w(\vfi(\xi)),\partial_{\xi_i}\vfi(\xi))>=0,
\eeq
we obtain that
\beq\label{eqcourb1}
\det(<w'(I)\partial_{\xi_i}\vfi(0),\partial_{\xi_j}\vfi(0)>)_{1\leq i,j\leq n-1}=(-1)^{n-1}\det(<w(I),\partial_{\xi_i}\partial_{\xi_j}\vfi(0)>)_{1\leq i,j\leq n-1}.
\eeq
Let $\alpha:=w'(I)^{-1}(w(I))$. First suppose that $\alpha\in T_I\nrja$. Then there exist real numbers $\mu_1,\dots \mu_{n-1}$ such that
$$\alpha=\sum_{j=1}^{n-1}\mu_j \partial_{\xi_j}\vfi(0),$$
i.e.
$$w(I)=\sum_{j=1}^{n-1}\mu_j w'(I)(\partial_{\xi_j}\vfi(0)).$$
Taking the bracket of the last equation with $\partial_{\xi_i}\vfi(0)$, we obtain that $K(I)=0$ via (\ref{eqcourb1}), and equality (\ref{courbformule}) holds since in this case $<w'(I)^{-1}(w(I)),w(I)>=0$.\

Suppose now that $\alpha\notin \nrja$. Then $\e:=(\alpha,\partial_{\xi_1}\vfi(0),\dots,\partial_{\xi_{n-1}}\vfi(0))$ is a basis of $\rd$. In view of equation (\ref{eqorth}), so is $\e':=(w(I),\partial_{\xi_1}\vfi(0),\dots,\partial_{\xi_{n-1}}\vfi(0))$. We write
\beq\label{eqcourb2}
\det(w'(I))=\det(w'(I))_{\e,\e'}\det(Id)_{\e',\e}.
\eeq
We easily get $\det(Id)_{\e',\e}=\frac{|w(I)^2|}{<w'(I)^{-1}(w(I)),w(I)>}$. Moreover, if $\alpha_{i,j}$ are real numbers such that 
$$w'(I)(\partial_{\xi_i}\vfi(0))=\sum_{r=1}^{n-1}\alpha_{i,r}\partial_{\xi_r}\vfi(0) +\alpha_{i,n} w(I),$$
then $\det(w'(I))_{\e,\e'}=\det(\alpha_{i,j})_{1\leq i,j\leq n-1}$. Tacking the bracket of the last equation with $\partial_{\xi_j}\vfi(0)$, we obtain that
$$\det(<w'(I)\partial_{\xi_i}\vfi(0),\partial_{\xi_j}\vfi(0)>)=\det(w'(I))_{\e,\e'}\det(<\partial_{\xi_i}\vfi(0),\partial_{\xi_j}\vfi(0)>)_{1\leq i,j\leq n-1}.$$
Together with equations (\ref{eqcourb1}), and (\ref{eqcourb2}), this proves the Lemma.
\end{proof}
\subsection{The Berry-Tabor formula}
Our goal here is to use Theorem \ref{gutztheo2} under preceeding hypotheses to give a STF in the integrable case involving geometric quantities.
The following lemma shows that assumptions of this theorem are satisfied.
\begin{lem}\label{intnorm}Make hypotheses (\ref{integrable}), (\ref{connectedness}), (\ref{globalAA}).
Let $z\in \nrj$, and denote by $(\theta,I)=\psi(z)$. Suppose that $\Phi_T(z)=z$ where $T\in\rr^*$. Then
\beq\label{eqinvertible}
w'(I) \mbox{ is invertible}\iff \ker(M_z(T)-Id)=\vect(J\nabla F_1(z),\dots,J\nabla F_n(z)).
\eeq
\beq
<w'(I)^{-1}(w(I)),w(I)>\neq 0 \iff \jgrad\notin (M_z(T)-Id)(T_z\nrj).
\eeq
Suppose moreover (\ref{NDint}) and (\ref{isoint}). Then periodic orbits of $\nrj\times\supp\hat{f}$ are $\nrj$-normal. Besides, if $(T,z)$ is such a periodic point, then we have $E_1(T,z)=\ker(M_z(T)-Id)^2=\rdd.$
\end{lem}
Note that Lemma \ref{toresisoles} actually follows from Proposition \ref{normalFP}.
\begin{proof}
Let us denote by 
\beq\label{W(z)}
W(z):=\vect(\nabla F_1(z),\dots,\nabla F_n(z)).
\eeq
For any integrable system of the form (\ref{integrable}), we have $E_1(T,z)=\rdd$. To see this, just apply Lemma \ref{Meyer} with $V=W=W(z)$. We introduce the canonical projection
\beq\label{canon}
u:\rd\to \mathbb{T}^n.
\eeq
Of course, the differential of $u$ at any point of $\rd$ is an isomorphism. Note that in (\ref{SDint}), to be rigorous, we should say $\dot{\theta}_t(\theta,I)=d_{\hat{\theta}}u(\nabla \hta(I))$, where  $\hat{\theta}\in\rd$ is such that $u(\hat{\theta})=\theta$.
If $(\theta,I):=\psi(z)$, we denote by $\widetilde{M}_{(\theta,I)}(t):=\vfi_T'(\theta,I)$. Then for all $(\tau,x)\in T_\theta \tore\times \rd$ we have
\beq\label{mtilde}
(\widetilde{M}_{(\theta,I)}(T) -Id)(\tau,x)=T(d_{\hat{\theta}}u(\hta''(I)x),0).
\eeq
We can read on this last formula that $T_\theta\tore \times \{ 0 \}\subset \ker(\widetilde{M}_{(\theta,I)}(T) -Id))$ and that $\hta''(I)$ is invertible if and only if $\dim(\ker(\widetilde{M}_{(\theta,I)}(T) -Id))=n$.\

Using properties of $\psi$, we note that
\beq\label{psijgrad}
\psi(\nrj)=\tore\times\nrja, \qquad d_z\psi(\jgrad)=(d_{\hat{\theta}}u(\nabla \hta(I)),0).
\eeq
Thus, in view of (\ref{mtilde}), $\jgrad\in (M_z(T)-Id)(T_z\nrj)$ if and only if there exists $(\tau,x)$ in $T_\theta \tore\times (\rr\nabla \hta(I))^\perp$ such that $(d_{\hat{\theta}}u(\nabla \hta(I)),0)=(\widetilde{M}_{(\theta,I)}(T) -Id)(\tau,x)=T(d_{\hat{\theta}}u( \hta''(I)x),0)$, i.e. if $\nabla \hta(I)=T\hta''(I)x$, where $x\perp \nabla \hta (I)$. That exactly means $<w'(I)^{-1}(w(I)),w(I)>= 0$.\

Since $M_z(T)$ is symplectic we have
\beq\label{eqkernel}
\ima(M_z(T)-I)=\ker(^tM_z(T)-Id)^\perp=J[\ker(M_z(T)-Id)]^\perp.
\eeq
We always have $JW(z)\subset \ker(M_z(T)-Id)$. Now if we make assumption (\ref{NDint}), in view of (\ref{eqinvertible}), we have equality. Moreover due to the involutions, we have $W(z)\subset JW(z)^\perp$, and in fact $W(z)= JW(z)^\perp$ for dimensional reasons. Coming back to (\ref{eqkernel}), we obtain $\ima(M_z(T)-I)=JW(z)=\ker(M_z(T)-Id)$, which means precisely $(M_z(T)-Id)^2=0$.
\end{proof}
We can now give a rigorous mathematical proof of the Berry-Tabor trace formula (see original papers \cite{BT1} and \cite{BT2}).
\begin{theo}\label{STFberrytabor}
Make hypotheses (\ref{integrable}), (\ref{connectedness}), (\ref{globalAA}), (\ref{NDint}) and (\ref{isoint}). Suppose for clarity that $0\notin\supp\hat{f}$. Then
$$\mathcal{G}_E(h)=\psi(E)\sum_{\substack{(I,T)\in\nrja\times \mbox{{\scriptsize supp}}\hat{f}:\\ \mathbb{M}:=\frac{Tw(I)}{2\pi}\in\zz^n} }
\hat{f}(T) \frac{ e^{\frac{i}{h} 2\pi <\mathbb{M};I>} e^{i\frac{\pi}{4}\beta_{I,T}} }{|w(I)|\, \sqrt{|K(I)|}\,|\mathbb{M}|^{\frac{n-1}{2}}}\; h^{\frac{1-n}{2}} 
+O(h^{\frac{1-n}{2}+1}),$$
where $\beta_{I,T}$ is an integer such that $e^{i\frac{\pi}{2} \beta_{I,T}}=i^{n-1}sign(K(I))$, $K(I)$ is the Gaussian curvature of $\nrja$ at $I\in\rd$ (cf (\ref{nrja}) and (\ref{curv})), and $w(I)$ is the frequency function on the torus $I$ (cf (\ref{frequence})).
\end{theo}
\begin{proof}
In view of Lemma \ref{intnorm} and Proposition \ref{normalFP}, we can apply Theorem \ref{gutztheo2}. Here, according to section \ref{intblabla}, connected components of $\ce$ are one to one with the set of $(T,I)\in \supp\hat{f}\times \nrja$ such that $Tw(I)\in (2\pi\zz)^n$. Each connected component has the dimension of a periodic torus, i.e. equal $n$. We obtain modulo $O(h^{\infty})$:
\beq\label{firststf}
\mathcal{G}_E(h)=(2\pi h)^{\frac{1-n}{2}}\psi(E)\sum_{\substack{(I,T)\in\nrja\times \mbox{{\scriptsize supp}}\hat{f}:\\ \mathbb{M}:=\frac{Tw(I)}{2\pi}\in\zz^n} }e^{\frac{i}{h}\A(T,I)}\hat{f}(T)\frac{1}{2\pi} 
\left( \int_{\Gamma_I}  d(T,z) d\sigma_I(z)+\sum_{j\geq 1}h^j a_{j,Y}\right),
\eeq
where $d\sigma_I$ is the Riemannian measure on the torus $\Gamma_I:=\psi^{-1}(\{(0,I)\})$ and $\A(T,I)$ is the constant value of the action on $\{T\}\times \Gamma_I$.\

First of all, we simplify the formula of the density. Let $(T,z)$ be a periodic point of $\supp \hat{f}\times\nrj$. Since $E_1=\rdd$, we have $\det(M_z(T) -Id)_{|_{V_1}} =1$ and $\det({w_0}_{|_{{E_1}}})=1$.  We introduce
$$\E_4(T,z):=\mathcal{E}_1(T,z)^{\perp}.$$
For sake of simplicity in the following ,we omit `$(T,z)$' in `$\E_j(T,z)$' notation.
First we note that we have $\mathcal{E}_4=J\E_1=J\E_2=W(z)$ (with the notation of (\ref{W(z)})).
We claim that
$$J(M_z(T)-Id)\E_4=\E_4.$$
Indeed we have seen in the proof of Lemma \ref{intnorm} that $(M_z(T)-Id)(\rdd)=JW(z)$ and $\rdd=\E_1+\E_4$ with $\E_1=\ker(M_z(T)-Id)$.
We note that since $\E_5=\E_4\cap T_z\nrj$, we have
$$\E_5\overset{\perp}{\oplus}\rr\grad=\E_4.$$
Let $\e_1\in \E_4$ such that $(M_z(T)-Id)\e_1=\jgrad$. Let $\beta''$ be an orthonormal basis of $\E_5$. Then, since $\jgrad\notin (M_z(T)-Id)(T_z\nrj)$, $\beta:=(\e_1,\beta'')$ is a basis of $\E_4$. Let $\beta':=(\grad,\beta'')$. It is of course also a basis of $\E_4$. Then
$$\det[J(M_z(T)-Id)_{|_{\E_4}}]=\det[J(M_z(T)-Id)_{|_{\E_4}}]_{\beta,\beta'}\det[Id]_{\beta',\beta}.$$
Since $\grad\perp\beta''$, we have $\det[J(M_z(T)-Id)_{|_{\E_4}}]_{\beta,\beta'}=-\det(\Pi_{\E_5}J(M_z(T)-Id)_{|_{\E_5}})$ and $\det[Id]_{\beta',\beta}=\norm{\grad}^2/<\e_1,\grad>$. Thus
$$\norm{\grad}^2\det(\Pi_{\E_5}J(M_z(T)-Id)_{|_{\E_5}})=-<\e_1,\grad>\det[(M_z(T)-Id)_{\E_4}].$$
Therefore, in view of Theorem \ref{gutztheo2},
\beq\label{densityint}
d(T,z)^2=\frac{(-1)^{n+1}i^{-(n+1)}}{ <\e_1(T,z),\grad> \det(J(M_z(T)-I)_{|_{\E_4}} ) }
\eeq
where $\e_1(T,z)$ is the element of $\E_4$ such that $(M_z(T)-Id)[\e_1(T,z)]=\jgrad$.
Let $\alpha=(\theta, I):=\psi(z)$ and $\widetilde{M}_\alpha(T):=d_z\psi(M_z(T))d_z\psi^{-1}$. As $\E_4=J\E_1$, and as $d_z\psi (\E_1)=T_\theta\tore\times \{0\}$, we have
$$\det(J(M_z(T)-I)_{|_{\E_4}} )=\det((M_z(T)-I)J_{|_{\E_1}} )=\det[(\widetilde{M}_\alpha(T)-Id)(d_z\psi J d_z\psi^{-1})_{|_{T_\theta\tore\times \{0\}}}].$$
Using notations of (\ref{canon}) we introduce $G_\theta: \{0\}\times\rd\to T_\theta\tore\times \{0\}$ defined by $$G_\theta(0,x):=(d_{\hat{\theta}} u(x),0).$$
We have
\beq\label{detE_4}
\det(J(M_z(T)-I)_{|_{\E_4}} )=\det[G_\theta^{-1} (\widetilde{M}_\alpha(T)-Id)_{|_{ \{0\}\times \rd}}]\det[(d_z\psi J d_z\psi^{-1}G_\theta)_{|_{ \{0\}\times \rd}}].
\eeq
We introduce $i:\{0\}\times\rd\to \rd$ defined by $i(0,x):=x$. Of course,
$$\det[G_\theta^{-1} (\widetilde{M}_\alpha(T)-Id)_{|_{ \{0\}\times \rd}}]=\det[i G_\theta^{-1} (\widetilde{M}_\alpha(T)-Id)i^{-1}].$$
In view of (\ref{mtilde}), for all $i G_\theta^{-1} (\widetilde{M}_\alpha(T)-Id)i^{-1}=T\hta''(I)$. Therefore we have
\beq\label{T1}
\det[G_\theta^{-1} (\widetilde{M}_\alpha(T)-Id)_{|_{ \{0\}\times \rd}}]=T^n \det(\hta''(I)).
\eeq
We obtain a parametrisation of $\Gamma_I=\psi^{-1}((0,I))$ with the following fonction $\vfi:[0,2\pi]^n\to \Gamma_I$ defined by
\beq\label{fifi}
\vfi(\tau):=\psi^{-1}(u(\tau),I), \quad \forall \tau\in [0,2\pi]^n.
\eeq
Differentiating the last equation leads to
$$d_{\vfi(\tau)} \psi \circ d_\tau\vfi\circ i=G_\theta.$$
Thus
\beq\label{coco}
\det[(d_{\vfi(\tau)}\psi \circ J \circ d_{\vfi(\tau)}\psi^{-1}\circ G_\theta)_{|_{ \{0\}\times \rd}}]=\det[(d_{\vfi(\tau)}\psi\circ J\circ d_\tau\vfi\circ i)_{|_{ \{0\}\times \rd}}]
=\det[( i \circ d_{\vfi(\tau)}\psi \circ J d_\tau\vfi )_{|_{\rd}}].
\eeq
We claim that we have $i\circ d_{\vfi(\tau)}\psi= ^td_\tau\vfi\circ J$ on $\E_4$. To check this, let $\beta\in\E_4$ and $y\in\rd$. By definition of $^td_\tau\vfi$, we have
$$<^td_\tau\vfi\circ J\beta,y>=<J\beta,d_\tau \vfi(y)> =
{\tilde{w}}_{\psi(\vfi(\tau))}(d_{\vfi(\tau)}\psi(\beta),d_{\vfi(\tau)}\psi\circ d_\tau\vfi(y)).$$
where $\tilde{w}$ is the symplectic form on $\tore\times\rd$. We have used here that $\psi$ is a symplectic map. Now, in view of (\ref{fifi}), and using the definition of $\tilde{w}$, the last term of the preceeding equation is equal to
$${\tilde{w}}_{\psi(\vfi(\tau))}((d_{\vfi(\tau)}\psi(\beta),(d_\tau u(y),0))=<i\circ d_{\vfi(\tau)}\psi(\beta),y>.$$
Thus we have proven that $i\circ d_{\vfi(\tau)}\psi= ^td_\tau\vfi\circ J$ on $\E_4$. Coming back to (\ref{coco}), we obtain
$$ \det[(d_{\vfi(\tau)}\psi \circ J \circ d_{\vfi(\tau)}\psi^{-1}\circ G_\theta)_{|_{ \{0\}\times \rd}}]=(-1)^n\det( ^td_\tau\vfi d_\tau\vfi)=:(-1)^n g_{\vfi}(\tau).$$
In view of (\ref{detE_4}) and (\ref{T1}), we have, if $z=\vfi(\tau)$
\beq\label{detE4}
\det(J(M_z(T)-I)_{|_{\E_4}} )=T^n \det(\hta''(I))g_{\vfi}(\tau)(-1)^n.
\eeq
We claim that 
\beq\label{e1grad}
<\e_1,\grad>=\frac{1}{T}<\hta''(I)^{-1}\nabla \hta(I),\nabla \hta(I)>.
\eeq
Indeed by definition of $\e_1$,  we have in view of (\ref{psijgrad})
$$(\tilde{M}_z(T)-Id)d_z\psi(\e_1)=(d_{\hat{\theta}}u(\nabla \hta(I)),0)$$
Let $(\tau_1,x_1):=d_z\psi(\e_1)$. Using (\ref{mtilde}) we obtain $Tx_1=\hta''(I)^{-1}\nabla \hta(I)$. Now, $\psi$ being a symplectic map, we have
$$<\e_1,\grad>=<J\e_1,\jgrad>={\tilde{w}}_{\psi(z)}(d_{z}\psi(\e_1),(d_{\hat{\theta}} u(\nabla\hta(I)),0))=<x_1,\nabla\hta(I)>,$$
which proves (\ref{e1grad}).
In view of (\ref{densityint}), (\ref{detE4}) and (\ref{e1grad}), we have obtained if $\vfi(\tau)=z$
$$d(T,z)^2=\frac{-Ti^{-(n+1)}}{<\hta''(I)^{-1}\nabla \hta(I),\nabla \hta(I)>T^n \det(\hta''(I))}\frac{1}{g_{\vfi}(\tau)}.$$
In view of Lemma \ref{courbur} we obtain
\beq\label{densityAA}
d(T,z)^2=(2\pi)^{-(n-1)}\frac{(-1)^ni^{-(n+1)}}{|w(I)|^2 K(I) |\mathbb{M}|^{n-1}}\frac{1}{g_{\vfi}(\tau)}.
\eeq
where $2\pi\mathbb{M}:=Tw(I)$. There exists an integrer $\beta=\beta(T,I)$ such that $e^{i\frac{\pi}{2} \beta}=i^{-(n+1)}\mbox{sign}(K(I))(-1)^n$ and that if $\vfi(\tau)=z$, then 
$$d(T,z)g_{\vfi}^{\ud}(\tau)=(2\pi)^{-(n-1)}\frac{e^{i\frac{\pi}{4}\beta}}{|w(I)| \sqrt{|K(I)|} |\mathbb{M}|^{\frac{n-1}{2}}}.$$
$$\frac{1}{2\pi}\int_{\Gamma_I}  d(T,z) d\sigma_I(z)=\frac{1}{2\pi}\int_{[0,2\pi]^n}d(T,\vfi(\tau))g_{\vfi}^{\ud}(\tau) d \tau=\frac{e^{i\frac{\pi}{4}\beta}}{|w(I)| \sqrt{|K(I)|} |\mathbb{M}|^{\frac{n-1}{2}}}.$$
Note that, since $z\mapsto d(T,z)$ is continuous, our integer $\beta$ is constant on the connected torus $\Gamma_I$. 
Endly, computing in action-angles coordinates gives 
$$\A(T,I)=2\pi <\mathbb{M};I>=T<w(I);I>.$$
This ends the proof of Theorem \ref{STFberrytabor}.
\end{proof}

\end{document}